\documentclass[fleqn,usenatbib]{mnras}
\usepackage[english]{babel}

\usepackage[T1]{fontenc}
\usepackage{ae,aecompl}

\usepackage{fancyhdr}
\usepackage{amsfonts}
\usepackage{amsmath}
\usepackage{amssymb}
\usepackage{multicol}
\usepackage{layout}
\usepackage{graphicx}
\usepackage{multirow}
\usepackage{hyperref}
\usepackage{color}
\usepackage{multirow}

\usepackage{times}
\usepackage{natbib}

\usepackage[outdir=./fig/]{epstopdf}

\newif\ifAMStwofonts
\AMStwofontstrue

\graphicspath{{./fig/}}

\title[HDE models]{Model selection and constraints from Holographic dark energy scenarios}

\date{Accepted ?, Received ?; in original form \today}
\author[I. A. Akhlaghi et~al.]{I. A. Akhlaghi $^1$, M. Malekjani$^{2}$  \thanks{malekjani@basu.ac.ir}, S. Basilakos$^{3}$ and H. Haghi $^1$
 \\
 $^1$ Department of Physics, Institute for advanced studies in Basic Sciences, Zanjan 45137-66731, Iran\\
$^2$ Department of Physics, Bu Ali Sina University, Hamedan 65178, Iran\\
$^3$ Academy of Athens, Research Center for Astronomy and Applied Mathematics, Soranou Efessiou 4, 11-527 Athens, Greec}

\pagerange{\pageref{firstpage}--\pageref{lastpage}} \pubyear{2017}

\begin{document}

\label{firstpage}

\maketitle
\begin{abstract}
In this study we combine the expansion and the growth data in order to 
investigate the ability of the three most popular holographic 
dark energy models, namely event future 
horizon, Ricci scale and Granda-Oliveros IR cutoffs, to fit the data.  
Using a standard $\chi^2$ minimization 
method we place tight constraints on 
the free parameters of the models. Based on 
the values of the Akaike and Bayesian information criteria  
we find that two out of three holographic dark energy models are disfavored 
by the data, because they predict a non-negligible 
amount of dark energy density at early enough times. Although  
the growth rate data are relatively consistent with the holographic dark energy 
models which are based on Ricci scale and Granda-Oliveros IR cutoffs,
the combined analysis provides strong 
indications against these models. Finally, we find that the model 
for which the holographic dark energy is related 
with the future horizon is  
consistent with the combined observational data.

\end{abstract}
\maketitle


\section{Introduction}
Since the discovery of the accelerated expansion of the universe 
in 1998 \citep{Riess1998,Perlmutter:1998np}, 
the role of dark energy (DE) in cosmic history has become one of the 
most complicated challenges in modern cosmology. 
Although the current cosmological data favor the Einstein 
cosmological constant model $\Lambda$ with the constant 
equation of state $w_{\rm \Lambda}=-1$ as 
the origin 
of the current accelerated expansion of the universe, this model suffers 
from two well known theoretical problems the so-called fine-tuning and cosmic coincidence issues \citep{Carroll2001,Peebles:2002gy,Padmanabhan2003,Copeland:2006wr,Frieman:2008sn,Li:2011wu,Bamba:2012vg}. In the last 
two decades, a large family of DE models with a 
time varying equation of state $w_{\rm de}(z)$ has been 
proposed to solve or at least to alleviate these problems. 
Unfortunately, in most of the cases the nature of DE 
is a big mystery in cosmology.  
The latter has given rise to some cosmologists to propose 
that the origin of DE is based on first principles, namely 
it is related with the effects of quantum 
gravity. Following this ideology one may consider that 
the holographic principle, which is one of the most fundamental
principle of quantum gravity, may play an important 
role towards solving the DE problem.

The holographic principle states that all information 
contained in a volume of space can be represented as a
hologram which corresponds to a theory locating on the 
boundary of that space \citep{tHooft1993,Susskind1995}.
In particular, according to the holographic principle, the number of degrees of 
freedom for a finite-size system is finite and bounded by the 
corresponding area of its boundary \citep{Cohen1999}. 
For a physical system with size $L$ the 
following relation is satisfied 
$L^3\rho_{\Lambda}\le LM^2_{\rm P}$, 
where $\rho_{\Lambda}$ is the  quantum zero-point energy density 
caused by the UV cutoff $\Lambda$ and 
$M_{\rm P}$ is the Planck mass ($M^2_{\rm P}=1/8\pi G$). 
In the context of cosmology, based on the holographic principle,  \cite{Li:2004rb}  proposed a new 
model of DE the so-called  holographic dark energy (HDE) model to interpret the positive acceleration of the universe.
The DE density in HDE models is given by \cite{Li:2004rb}
\begin{eqnarray}\label{eq:rho_hde1}
\rho_{\rm de}=3n^2M_{\rm P}^2L^{-2}\;,
\end{eqnarray}
where $n$ is a positive numerical constant. The important point is that the HDE model is defined in terms of the IR cutoff $L$. 
In the literature, there is an intense debate regarding scale of the IR cutoff. The basic cases are the following.\\
\begin{itemize}
\item{\bf Hubble horizon:}
The simplest choice is the Hubble length, i.e., $L=H^{-1}$. In fact in this case the holographic principle 
suggests that the energy density of DE is proportional to the square of the 
Hubble parameter, i.e., $\rho_{\rm de} \propto H^2$. 
In principle this choice solves the  fine-tuning problem, but the equation of state of DE is zero and thus the current accelerated 
expansion is impossible to take place
\citep{Horava2000,Cataldo2001,Thomas2002,Hsu2004}.\\

  \item{\bf Particle horizon:} if we select the particle horizon to be the IR cutoff then there is again a problem
 because it is impossible for this particular HDE model to provide an accelerated expansion of the universe \citep{Li:2004rb}.\\
  \item{\bf Future event horizon:} 
Here we choose $L$ to be the future event horizon \cite{Li:2004rb} which is given by
  \begin{equation}\label{eq:h2}
  R_{\rm h}=a \int^{\infty}_t{\frac{dt}{a(t)}}=a\int_a^{\infty}{\frac{da}{Ha^2(t)}}\;,
  \end{equation}
  where $a$ is the scale factor, $H$ is the Hubble parameter and $t$ is the cosmic time. In this case the DE 
energy density is written as 
   \begin{equation}\label{eq:rho_hde2}
   \rho_{\rm de}=3n^2M_{\rm P}^2R_{\rm h}^{-2}\;.
   \end{equation}
   It has been found that the current 
HDE model accommodates the late time acceleration and it is 
consistent with the cosmological observations \citep{Pavon2005,Zimdahl2007}. 
Also, the coincidence and the fine-tuning problems are typically 
alleviated at this length scale \citep{Li:2004rb}. The HDE model with the event horizon IR cutoff has been widely studied and constrained using cosmological data \citep{Huang:2004wt,Kao:2005xp,Zhang:2005hs,Wang:2005ph,Chang:2005ph,Zhang:2007sh,Micheletti:2009jy,Xu:2012aw,Zhang:2013mca,Li:2013dha,Zhang:2014ija,Zhang:2015rha}.\\
  \item{\bf Ricci scale cutoff :} In this model, the IR scale of the universe is the curvature of spacetime, namely the Ricci scalar 
\citep{Nojiri:2005pu,Gao:2007ep,Zhang:2009un}. For a spatially 
flat FRW universe, the Ricci scalar reads
  $R=-6(\dot{H}+H^2)$ which implies that the DE energy density becomes 
$\rho_{\rm de}=-(\kappa/16\pi)R$, where $\kappa$ is a numerical constant. 
Considering $R\sim L^{-2}$, the DE energy density of Ricci HDE model is given by
  \begin{equation}\label{eq:rho_Ricci1}
  \rho_{\rm de}=\frac{3\kappa}{8\pi}(\dot{H}+2H^2).
  \end{equation}
  It has been found that the Ricci HDE model is consistent
with the supernova type Ia data \citep{Zhang:2009un,Easson:2010av}. 

  \item{\bf Granda \& Oliveros (GO) cutoff:} As we have already mentioned above, 
the Hubble scale alone cannot justify the current acceleration of the 
universe and therefore it cannot be considered as an IR cutoff for HDE models. 
The simplest generalization that produces cosmic acceleration 
is to combine the Hubble parameter together with its time derivative \citep[see][]{Granda:2008dk}. 
In this case the energy density of DE takes the form
  \begin{eqnarray}\label{eq:rho_GO}
\rho_{\rm de}=3 (\alpha H^2+\beta \dot{H})\;,
  \end{eqnarray}
where $\alpha$ and $\beta$ are the numerical constants of the model .
Notice, that similar considerations regarding the functional form of 
Eq.(\ref{eq:rho_GO}) can be found in \citep[][]{Easson:2010av,Easson:2010xf,Basilakos:2014tha}.
Obviously, the DE density (\ref{eq:rho_Ricci1}) can be seen as a particular case 
of Eq.(\ref{eq:rho_GO}).  

\end{itemize}

In this work we attempt to test the performance of the most popular 
HDE models against the latest cosmological data.
Notice that in the current study we decide to ignore those HDE models 
for which the particle horizon IR cutoff is equal to $H^{-1}$, since 
they do not recover the correct 
equation of state for DE. 
In addition to background evolution, we also explore the HDE 
models at the perturbation level using 
the growth rate of large scale structures in the linear regime
\citep{Tegmark:2003ud}.
It is well known that DE not only accelerates the expansion  
of the universe but also it affects the growth of matter perturbations. 
Interestingly, in the context HDE models, for which the EoS parameter
varies with time, one can consider that DE clumps 
in a similar fashion to dark matter \citep{Abramo2007,Abramo:2008ip,Batista:2013oca,Batista:2014uoa,ArmendarizPicon:1999rj,ArmendarizPicon:2000dh,Mehrabi:2014ema,Mehrabi:2015hva,Mehrabi:2015kta,Malekjani:2016edh}. 
Indeed, the key quantity that describes the clustering of DE is 
the so called effective sound speed $c^2_{\rm eff}=\delta p_{\rm de}/\delta \rho_{\rm de}$. Specifically, in the case 
of $c^2_{\rm eff}=1$ (in units of the speed of light), the 
sound horizon of DE is larger than the Hubble length 
which implies that DE perturbations inside the Hubble scale 
cannot grow (homogeneous DE models). 
On the other hand, for $c^2_{\rm eff}=0$ the sound horizon 
is quite small with respect to 
the Hubble radius and thus the fluctuations of DE can grow 
due to gravitational instability in a similar fashion 
to matter perturbation \citep{ArmendarizPicon:1999rj,ArmendarizPicon:2000dh,Garriga:1999vw,Akhoury:2011hr}.
Notice, that the clustered DE scenario 
has been extensively studied in the literature \citep{Erickson:2001bq,Bean:2003fb,Ballesteros:2008qk,dePutter:2010vy,Sapone:2012nh,Dossett:2013npa,Basse:2013zua,Batista:2013oca,Batista:2014uoa,Pace:2013pea,Pace:2014taa,Malekjani:2016edh,Mehrabi:2014ema,Mehrabi:2015hva,Mehrabi:2015kta,Nazari-Pooya:2016bra}. 
Although it is difficult to directly measure the amount of DE clustering, it has been shown that the clustered DE models fit the growth data equally well to 
homogeneous DE scenarios  \citep{Mehrabi:2014ema,Basilakos:2014yda,Mehrabi:2015kta,Malekjani:2016edh}.

In order to study DE at the background and perturbation levels, we need to 
set up a general formalism where the background 
geometrical data including SnIa, baryonic acoustic oscillation (BAO), cosmic microwave background (CMB) shift parameter, Hubble expansion $H(z)$, and
 big bang nucleosynthesis (BBN) are combined with the growth rate data, namely 
$f(z)\sigma_8$ \citep[for more details, see][]{Cooray:2003hd,Corasaniti:2005pq,Basilakos:2010fb,Blake:2011rj,Nesseris:2011pc,Basilakos:2012uu,Yang:2013hra,mota1,mota2,mota3,mota4,mota5,mota6,Contreras:2013bol,Chuang:2013hya,Li:2014mua,Basilakos:2014yda,Mehrabi:2015kta,Mehrabi:2015hva,Basilakos:2016xob,mota7,Malekjani:2016edh,Fay:2016yow,Rivera:2016zzr}.
In particular, \cite{Mehrabi:2015kta} studied the HDE model 
with future event horizon by applying an overall likelihood analysis 
using the Markov chain Monte Carlo (MCMC) technique 
in order to quantify the free parameters of the model. 
\cite{Mehrabi:2015kta} found that in the framework of the above HDE model
both clustered and homogeneous scenarios 
fit the observational data equally well with respect to that of 
the concordance $\Lambda$CDM model.

In this article we extend the work of \cite{Mehrabi:2015kta} 
to a more general case, namely the explored HDE models are considered 
with different IR cutoffs (see Table \ref{tab:table1}). 
We organize the paper as follows. 
In section (\ref{cosmology}) we present the main 
cosmological ingredients of the HDE models at the 
background and perturbation levels. 
In section (\ref{data_analysis}), we perform a
joint statistical analysis in order to 
place constraints on the free parameters of the HDE models using solely
expansion data (SnIa, BAO, CMB, $H(z)$ and BBN).  
Then, using the growth rate data 
we check the performance of the current HDE models at the perturbation level.
Finally, based on Akaike and Bayesian information criteria 
we study the ability of the combined (expansion+growth) data in 
constraining the cosmological parameters
of HDE models, including that of $\Lambda$CDM.
Finally, in section (\ref{conclusion}) we provide the 
conclusions of our study.

\begin{table}\label{tab:table1}
	\caption{Different HDE models based on various IR cutoffs considered in this work.}
	\begin{tabular}{c  c  c }
		\hline \hline
		Model (1): & HDE with event horizon IR cutoff\\
		\hline
		Model (2): & HDE with Ricci scale IR cutoff\\
		\hline
		Model (3): & HDE with GO IR cutoff\\
		\hline \hline
	\end{tabular}
\end{table}

\section{Cosmology in HDE models}\label{cosmology}
In this section we present the main elements of 
the HDE models introduced in Table (\ref{tab:table1}). In particular, we briefly present the main ingredients of the models
at the expansion and perturbation levels respectively.

\subsection{Background cosmology}\label{background-1}
 In the framework of spatially flat FRW metric if we consider that 
the universe is filled by radiation, pressure-less matter and DE then 
the Hubble parameter $H$ is given by
 \begin{equation}\label{eqn:fh3}
 H^{2}=\frac{1}{3M_{Pl}^{2}}(\rho_{\rm r}+\rho_{\rm m}+\rho_{\rm de})\;,
 \end{equation}
where $\rho_{\rm r}$, $\rho_{\rm m}$ and $\rho_{\rm de}$ are 
the corresponding densities of radiation, pressure-less matter and 
DE. In the case of a simple non-interacting system for 
which the cosmic fluids evolve separately, we can write the 
following continuity equations which describe the density evolution of  
each cosmic fluid

\begin{eqnarray}
&&\dot{\rho}_{\rm r}+4H\rho_{\rm r}=0, \label{eqn:contr}\\
&&\dot{\rho}_{\rm m}+3H\rho_{\rm m}=0, \label{eqn:contmt}\\
&&\dot{\rho}_{\rm de}+3H(1+w_{\rm de})\rho_{\rm de}=0\;,\label{eqn:contdt}
\end{eqnarray}
where the over dot is the derivative with respect to cosmic time and $w_{\rm de}$ is the EoS parameter of DE. Bellow, for the current HDE models 
we derive the functional form of the Hubble parameter.

\begin{itemize}
	\item{\bf Model 1:}
	Taking the time derivative of Eq.~(\ref{eqn:fh3}) 
using Eqs.~(\ref{eq:rho_hde1}), (\ref{eqn:contr},\ref{eqn:contmt} \ref{eqn:contdt}) and the relation $\dot{R}_{\rm h}=1+HR_{\rm h}$ the corresponding 
EoS parameter can be easily obtained as \citep{Li2004}
	\begin{equation}\label{eqn:eos}
	w_{\rm de}(z)=-\frac{1}{3}-\frac{2\sqrt{\Omega_{\rm de}(z)}}{3n}\;,
	\end{equation}
	where $\Omega_{\rm de}$ is the dimensionless density parameter of the DE component. Now, taking the time derivative of $\Omega_{\rm de}/\Omega_{\rm c}=1/(HR_{\rm h})^2$ and using the relation between redshift and scale factor $z=a^{-1}-1$, we can obtain the following  
differential equation 
	 \begin{equation}\label{eqn:difOmega_DE}
	 \frac{d\Omega_{\rm de}(z)}{dz}=\frac{3w_{\rm de}(z)\Omega_{\rm de}(z)[1-\Omega_{\rm de}(z)]}{(1+z)}\;.
	 \end{equation}
	 Also, using the Friedmann Eq.(\ref{eqn:fh3}) and the 
continuity Eqs. (\ref{eqn:contr},\ref{eqn:contmt} and \ref{eqn:contdt}), the dimensionless Hubble parameter $E(z)=H(z)/H_{0}$ of the current HDE model 
is written as
	 \begin{equation}\label{eq:hubb_dimless}
	 E^2(z)=\frac{\Omega_{\rm r0}(1+z)^4+\Omega_{\rm m0}(1+z)^3}{1-\Omega_{\rm de}(z)}\;,
	 \end{equation}
	 where $\Omega_{\rm r0}$ and $\Omega_{\rm m0}$ are the present values 
of the dimensionless densities, namely radiation and matter. 
Note that equations (\ref{eqn:eos}, \ref{eqn:difOmega_DE} \& \ref{eq:hubb_dimless}) form a system of equations a solution of which provides
the evolution of the main cosmological quantities 
$w_{\rm de}$, $\Omega_{\rm de}$ and $E(z)$.
Moreover, the free parameter $n$ plays an essential role 
in order to determine the cosmic evolution of DE in this model. 
Indeed, in the case of $n=1$, the EoS parameter 
asymptotically tends to $w_{\rm \Lambda}=-1$ in the far future. 
For $n>1$, the EoS parameter is always greater than $-1$ 
so the current HDE model behaves as a quintessence DE scenario.
On the other hand, if $n<1$ then the EoS parameter 
can cross the phantom line $w=-1$, leading to a 
phantom universe with a big-rip as its ultimate fate. 
Clearly, the latter discussion points that 
it is crucial to constrain the value of $n$. 
\end{itemize}

\begin{itemize}
\item{\bf Model 2:}
	Inserting Eq.(\ref{eq:rho_Ricci1}) into Friedmann 
Eq.~(\ref{eqn:fh3}), we can obtain the following equation 
of the dimensionless Hubble parameter
	\begin{eqnarray}\label{eq:hubb_ricci}
	&&E^2(z)=\Omega_{\rm m0}(1+z)^3+\Omega_{\rm r0}(1+z)^4\nonumber\\
	&&+(\frac{\kappa}{2-\kappa})\Omega_{\rm m0}(1+z)^3+f_0(1+z)^{4-\frac{2}{\kappa}}\;	
	\end{eqnarray}
	where $f_0=1-\Omega_{\rm r0}-2\Omega_{\rm m0}/(2-\kappa)$.
	From Eq.(\ref{eq:hubb_ricci}), we may obtain 
	\begin{eqnarray}\label{eq:omega_de_model2}
	&&\Omega_{\rm de}(z)=(\frac{\kappa}{2-\kappa})\Omega_{\rm m0}(1+z)^3+f_0(1+z)^{4-\frac{2}{\kappa}}\;.
	\end{eqnarray}
It is interesting to mention that in the case of $\kappa=1/2$, model (2) 
reduces to cosmological constant $\Lambda$ plus the component of pressureless matter. We will show that the parameter $\kappa$ is not a 
free parameter but it is related to energy densities of radiation and matter ( see Eq.\ref{kappa}). Generally, based on Eq.(\ref{eqn:contdt}), the EoS parameter of DE reads 
	\begin{eqnarray}\label{eq:eos_model2}
	w_{\rm de}(z)=-1+\frac{1+z}{3}\frac{d\ln{\Omega_{\rm de}(z)}}{dz}\;,
	\end{eqnarray}
	where in this case the evolution of 
$\Omega_{\rm de}$ is given Eq.(\ref{eq:omega_de_model2}). 
Therefore, for the current HDE model we consider the coupled system of Eqs. (\ref{eq:hubb_ricci}, \ref{eq:omega_de_model2} \& \ref{eq:eos_model2}) in order 
to study the expansion history of the universe.
\end{itemize}

\begin{itemize}
	\item{\bf Model 3:}
	Substituting Eq.(\ref{eq:rho_GO}) into Friedmann  Eq.~(\ref{eqn:fh3})
and after some calculations we arrive at 
	\begin{eqnarray}\label{eq:hub_model3}
	E^2(z)=\Omega_{\rm r0}(1+z)^4+\Omega_{\rm m0}(1+z)^3+\Omega_{\rm de}(z),
	\end{eqnarray}
where the quantity $\Omega_{\rm de}(z)$ is given by
	\begin{eqnarray}\label{eq:omeg_DE_model3}
	&&\Omega_{\rm de}(z)=\frac{2\alpha-3\beta}{-2\alpha+3\beta+2}[\Omega_{\rm m0}(1+z)^3+\Omega_{\rm r0}(1+z)^4]\nonumber\\
	&&+[1-\frac{2}{-2\alpha+3\beta+2}(\Omega_{\rm m0}+\Omega_{\rm r0})](1+z)^{\frac{2(\alpha-1)}{\beta}} .
	\end{eqnarray}
	 Similar to model (2), we see that in the case of $\alpha=1$, 
model (3) contains a cosmological constant $\Lambda$ plus a pressure less matter. Notice, that the parameter $\alpha$ is not a free parameter, namely it can be 
fixed by the following relation \citep[see][]{Granda:2008dk} 
	 \begin{eqnarray}\label{alpha1}
	  \alpha=\frac{1}{2}\left(2\Omega_{\rm de0}+\beta[3\Omega_{\rm m0}+4\Omega_{\rm r0}+3(1+w_{\rm de}(z=0))\Omega_{\rm de0}]\right)\;,
	 \end{eqnarray}
	 where for a spatially flat universe we have
$\Omega_{\rm de0}=1-\Omega_{\rm m0}+\Omega_{\rm r0}$. 
Now, using Eqs. (\ref{eq:eos_model2}, \ref{eq:hub_model3} \& \ref{eq:omeg_DE_model3}) and after some calculations, we obtain the following relation 
	 \begin{eqnarray}\label{w0alpha}
	 w_{\rm de}(z=0)=\frac{2(\Omega_{\rm m0}+\Omega_{\rm r0})[1-\Omega_{\rm m0}-\Omega_{\rm r0}-\alpha]+3\beta}{3\beta(\Omega_{\rm m0}+\Omega_{\rm r0}-1)}\;.
	 \end{eqnarray}
	 Inserting Eq.(\ref{w0alpha}) into (\ref{alpha1}) 
the parameter $\alpha$ is written in terms of the other cosmological 
parameters as follows 
	 \begin{eqnarray}
	 \alpha=1-\Omega_{\rm r0}-\Omega_{\rm m0}.
	 \end{eqnarray}
We observe that for 
$\beta=\kappa$ and $\alpha=2\kappa$, model (3) boils down to 
model (2). Hence, the parameter $\kappa$ of model (2) is 
not a free parameter but it is given by 
	  \begin{eqnarray}\label{kappa}
	  \kappa=\frac{1-\Omega_{\rm r0}-\Omega_{\rm m0}}{2}\;.
	  \end{eqnarray}
\end{itemize}

\begin{figure*}
	\centering
	\includegraphics[width=8cm]{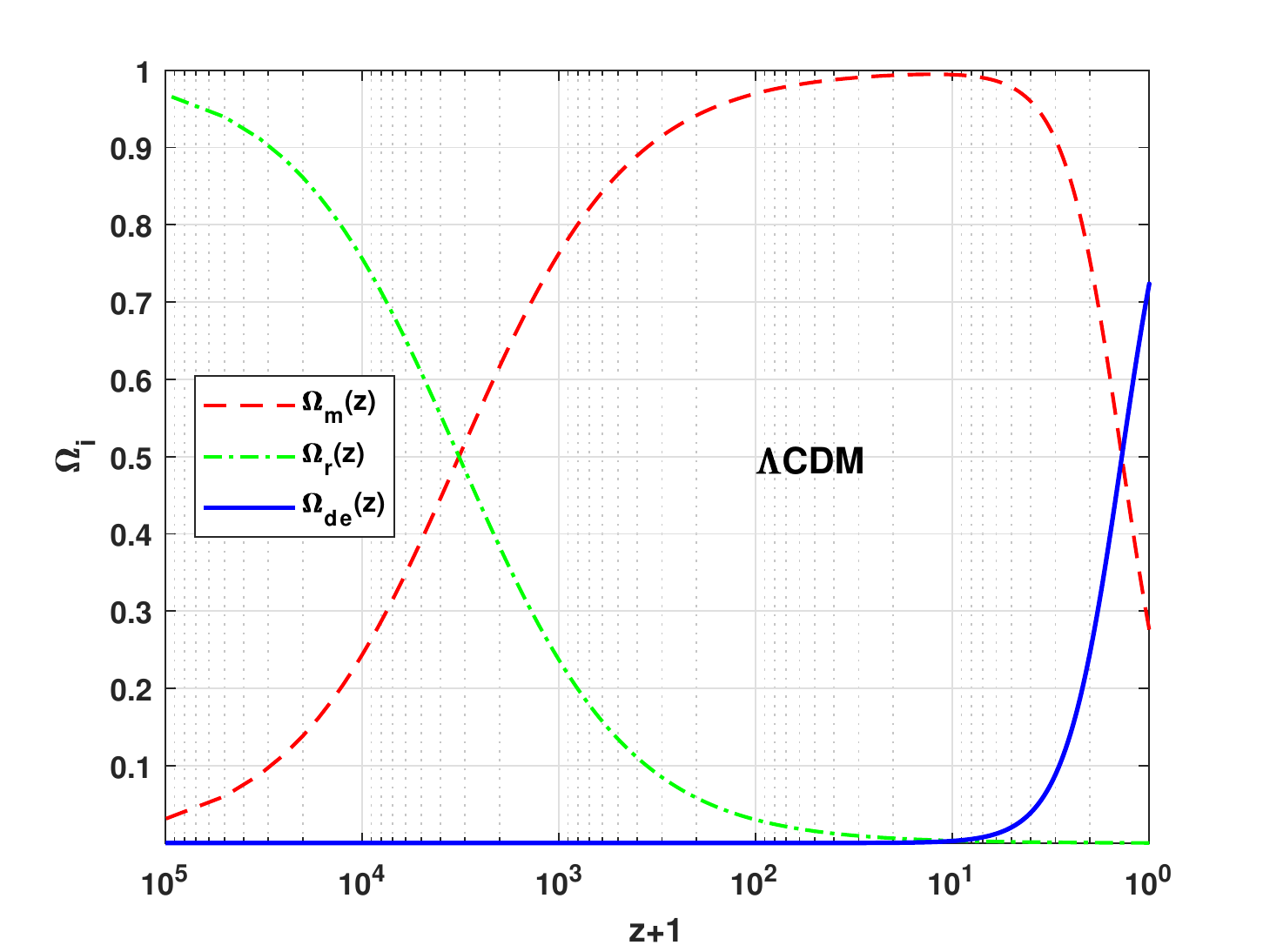}
	\includegraphics[width=8cm]{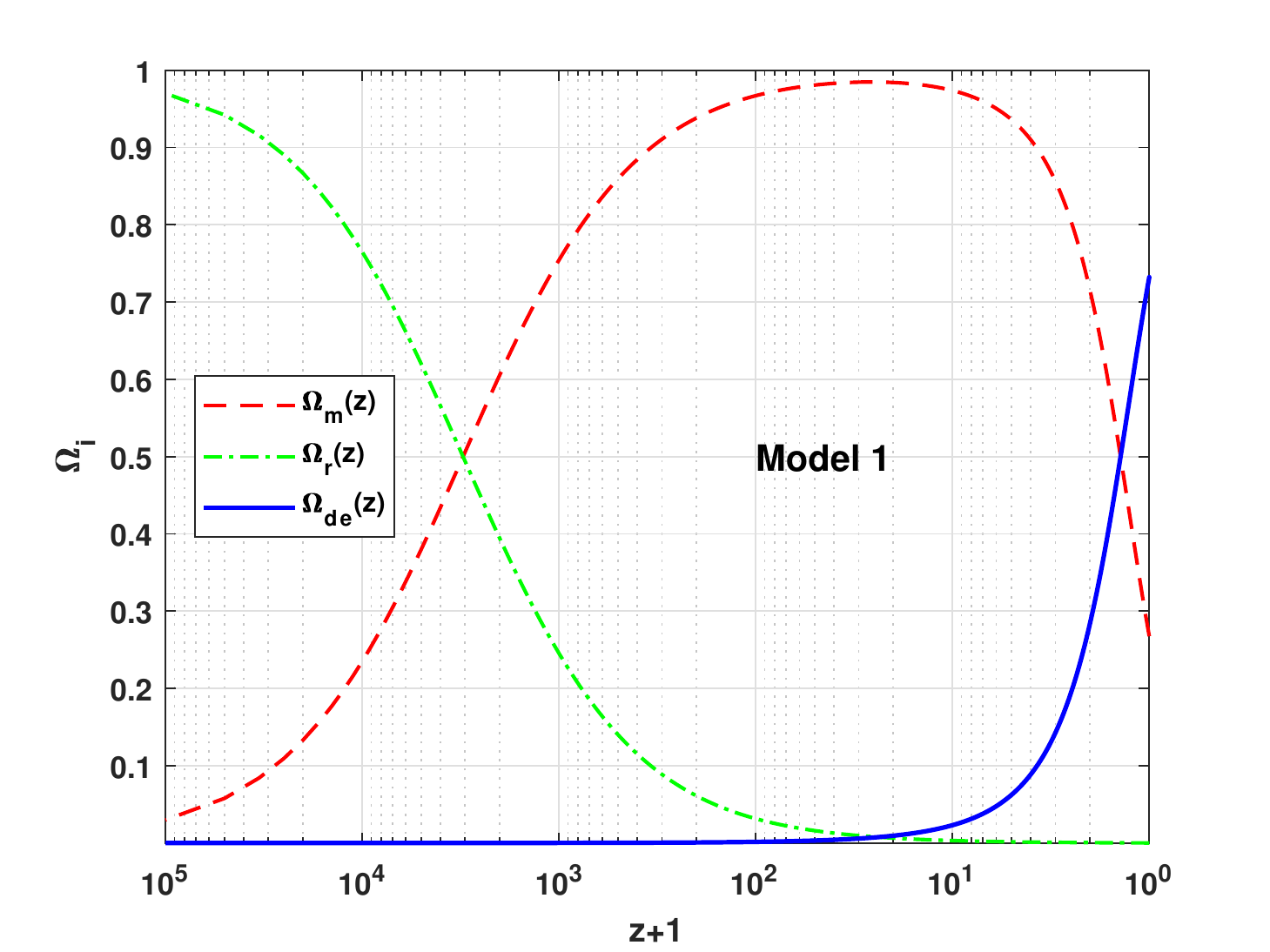}
	\includegraphics[width=8cm]{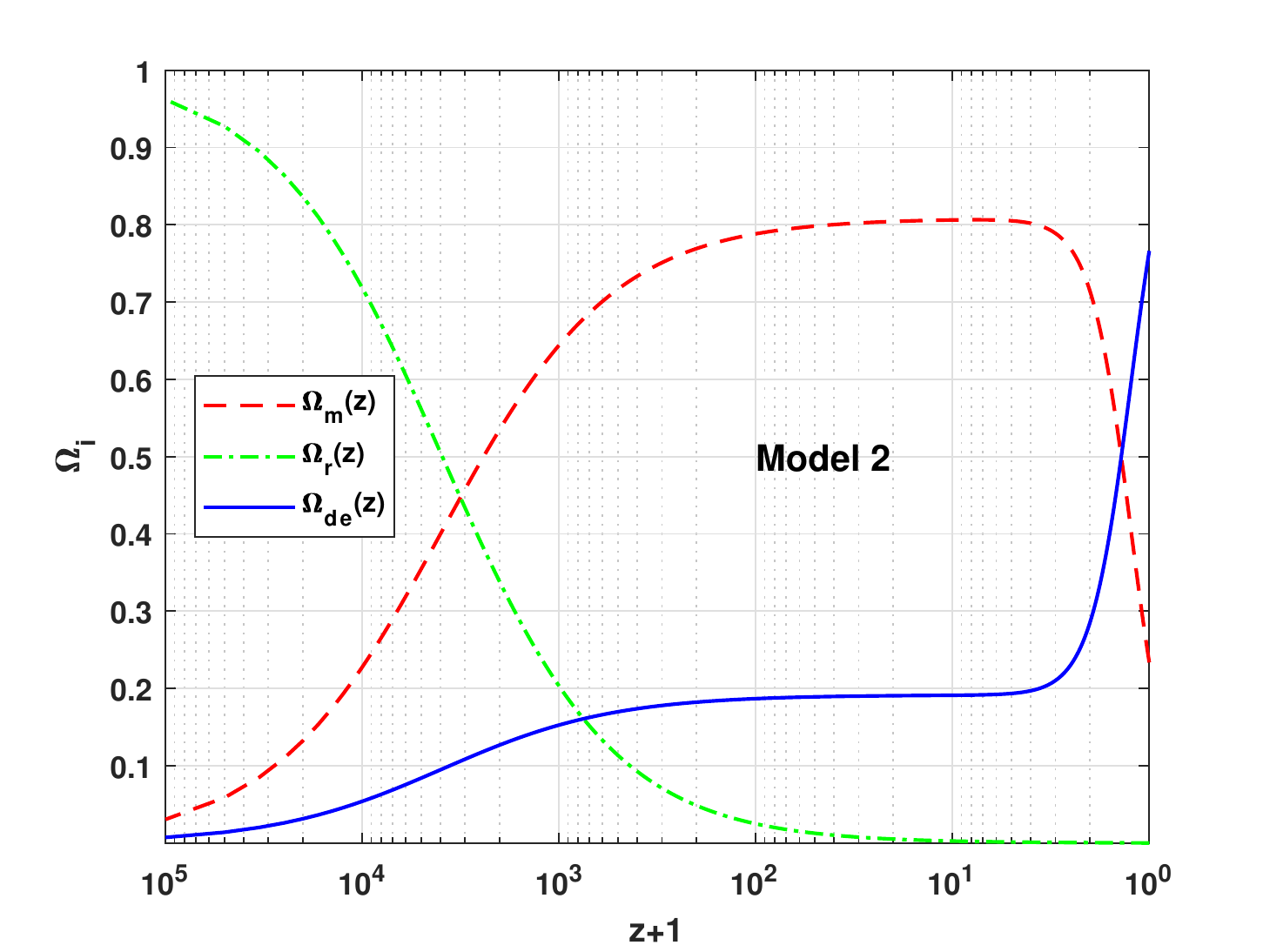}
	\includegraphics[width=8cm]{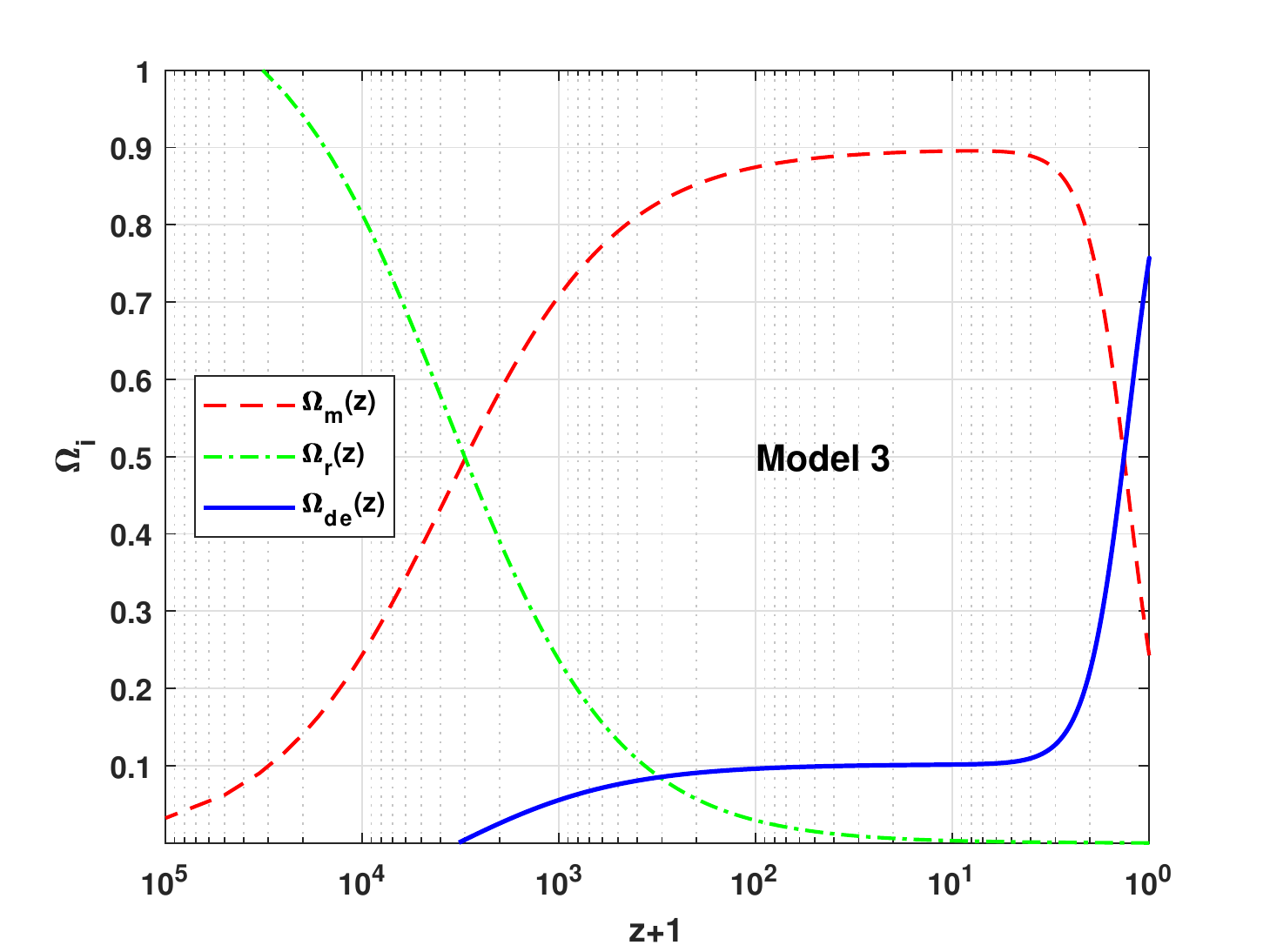}
	\caption{Redshift Evolution of radiation energy density $\Omega_{\rm r}(z)$, matter energy density $\Omega_{\rm m}(z)$ and DE density $\Omega_{\rm de}(z)$ 
		for different cosmological models explored in this work. 
		The style of the curves are presented in the inner panels.}
	\label{fig:energy_density}
\end{figure*}

\begin{figure*}
	\centering
	\includegraphics[width=8.5cm]{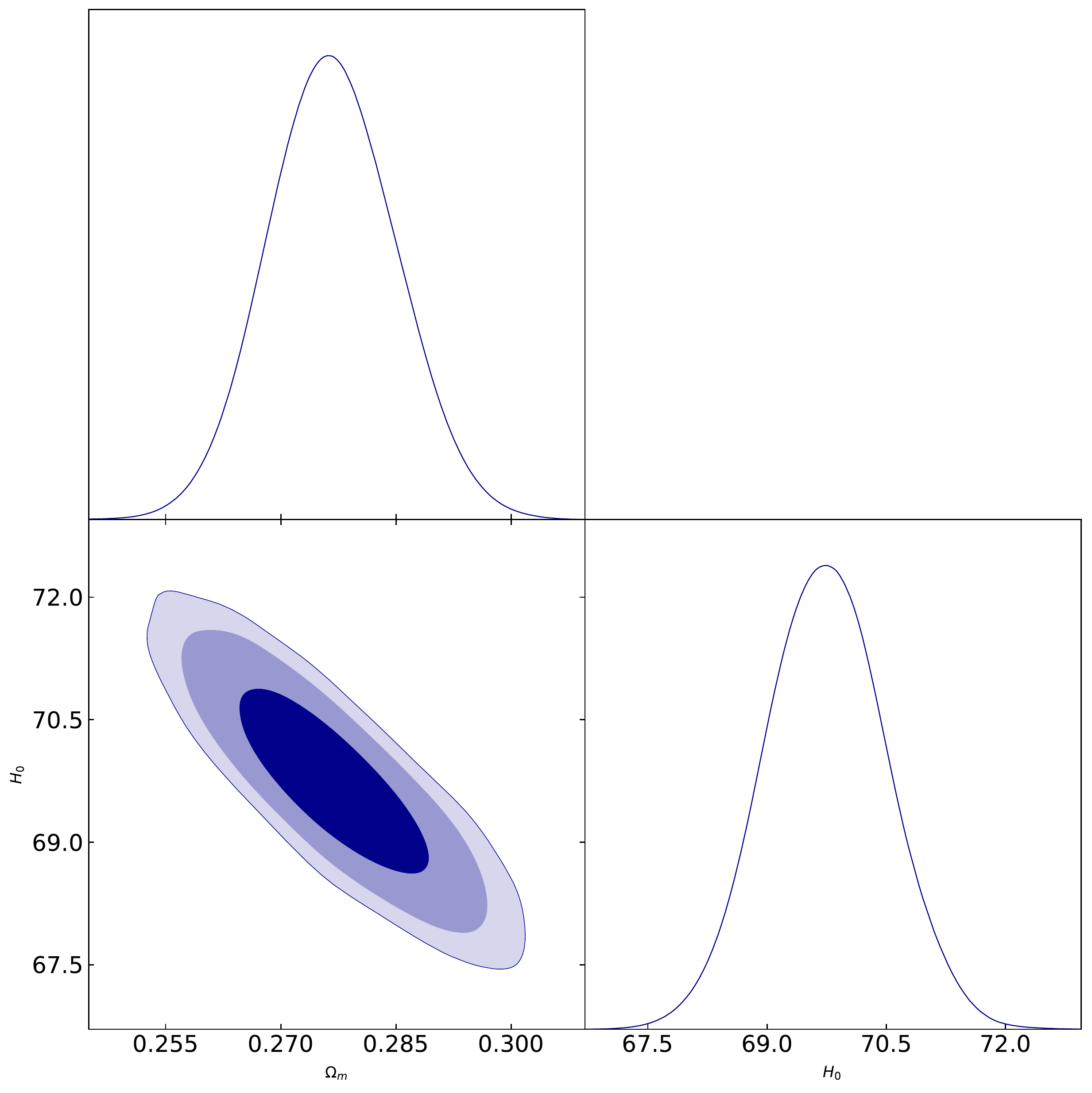}
	\includegraphics[width=8.5cm]{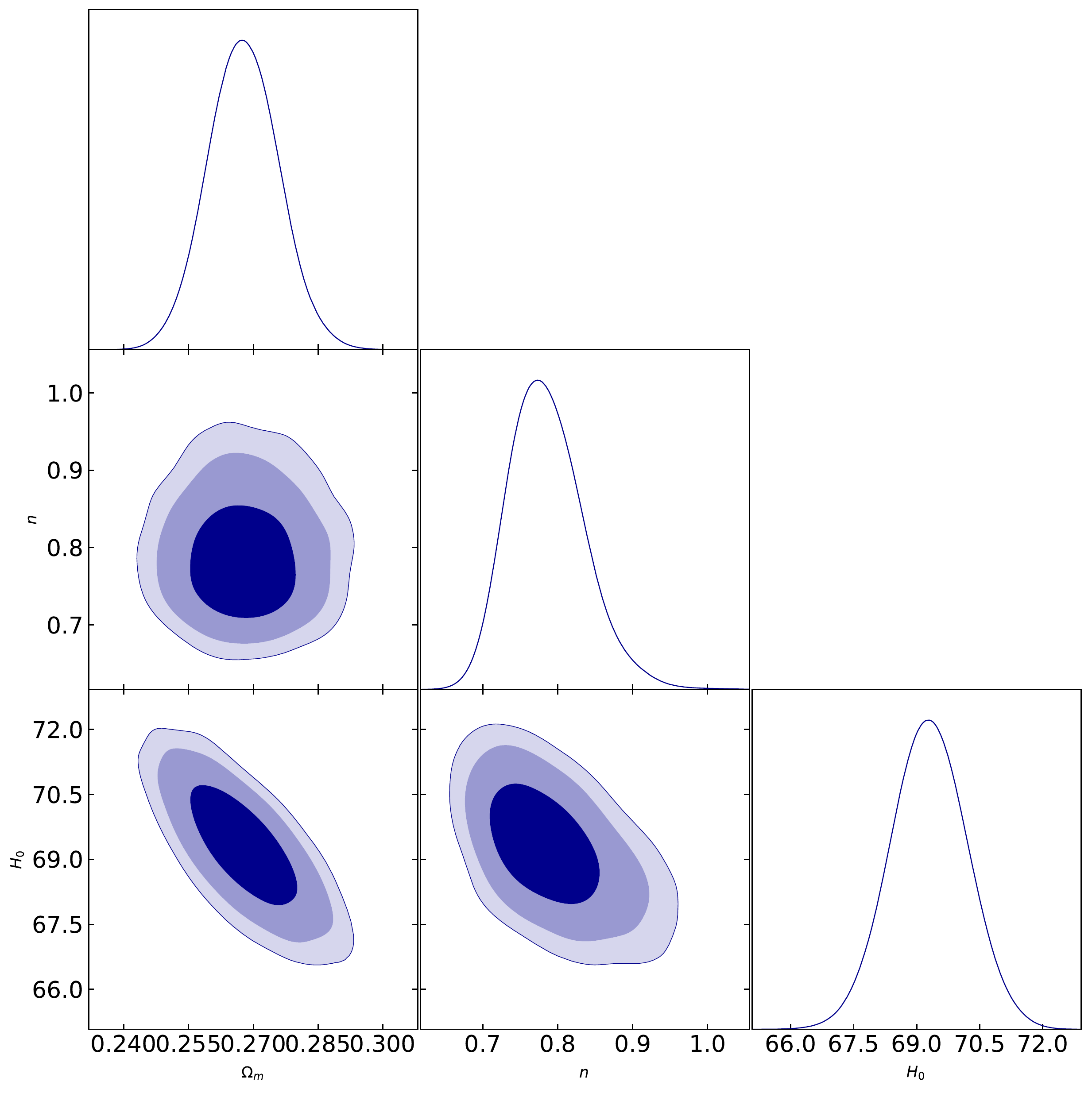}
	\includegraphics[width=8.5cm]{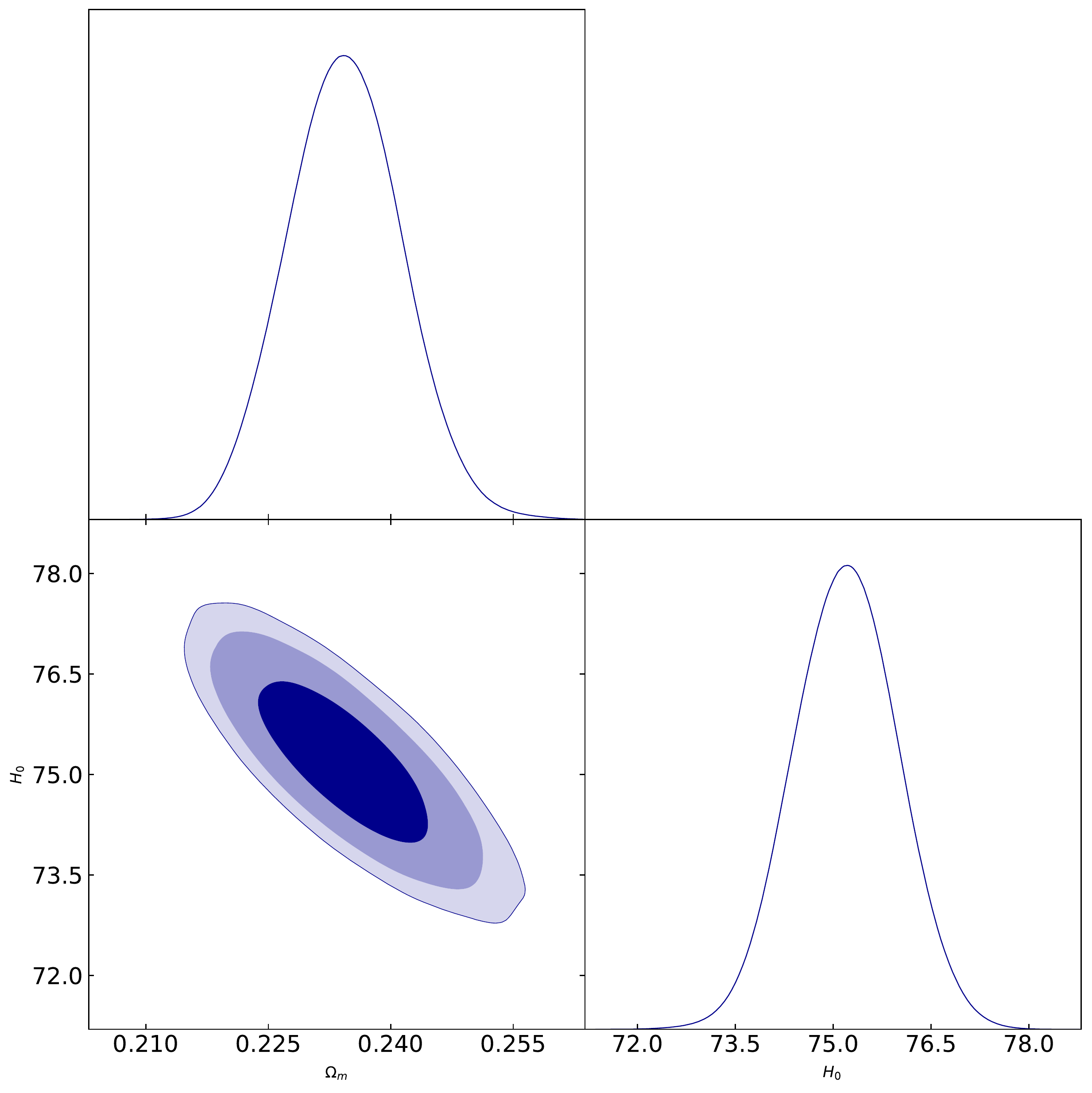}
	\includegraphics[width=8.5cm]{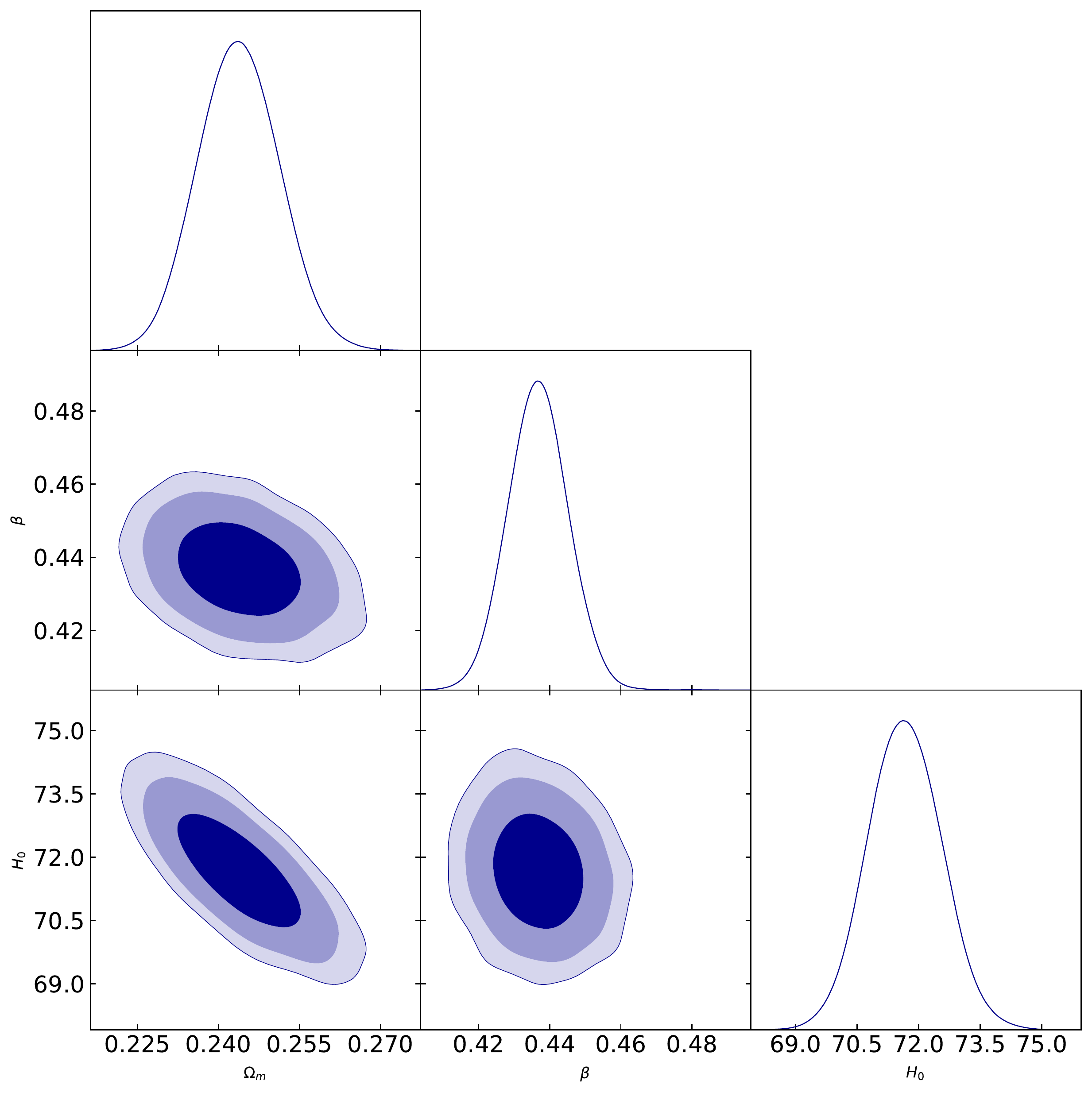}
	\caption{ The $1\sigma$,$2\sigma$ and $3\sigma$ contours 
		for various cosmological parameters using the background data. Different 
		panels correspond to $\Lambda$CDM model (up-left), HDE model 1 (up-right), 
		model 2 (bottom-left) and model 3 (bottom-right).}
	\label{fig:back}
\end{figure*}
\begin{figure*} 
	\centering
	\includegraphics[width=8cm]{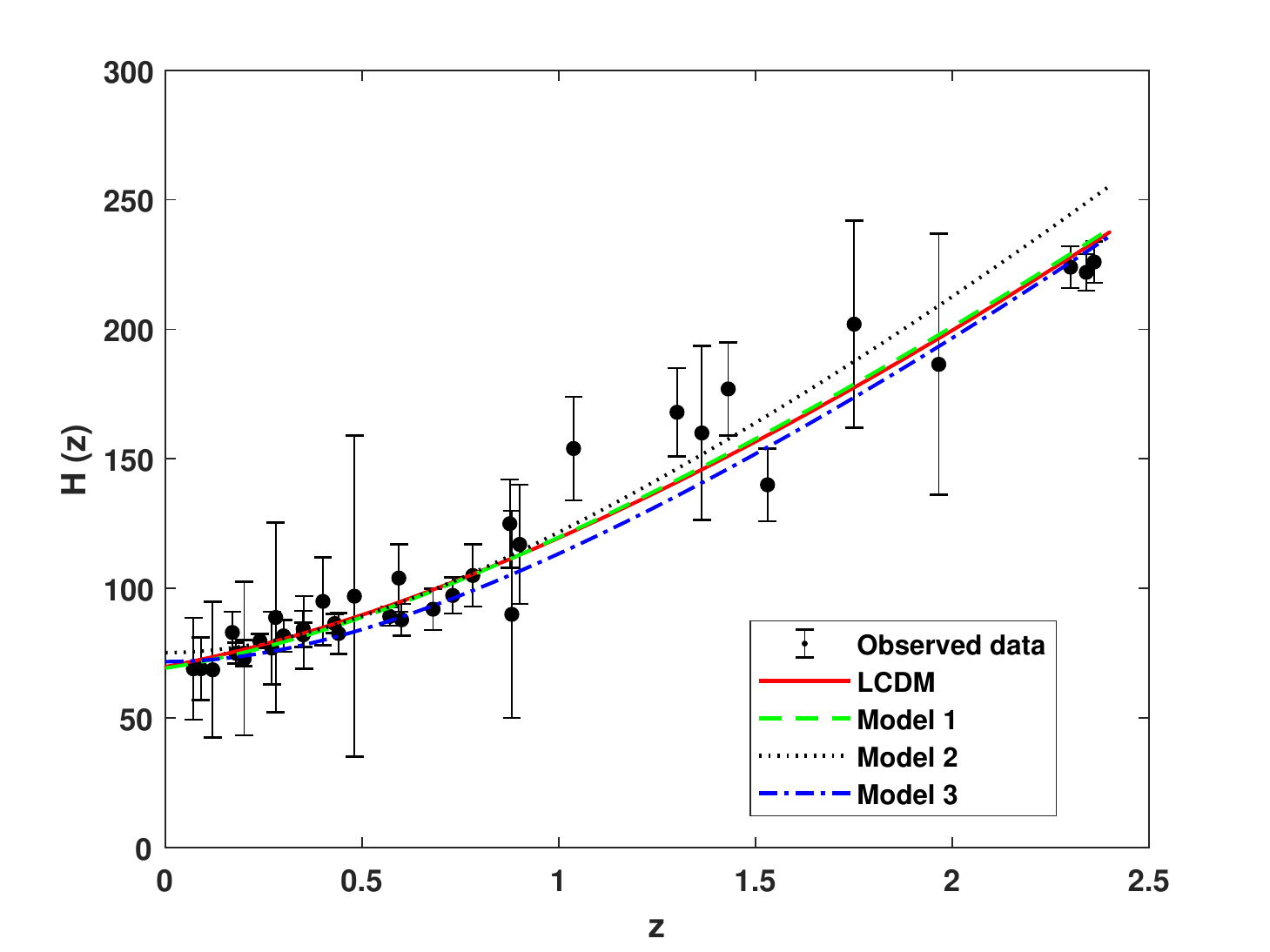}
	\includegraphics[width=8cm]{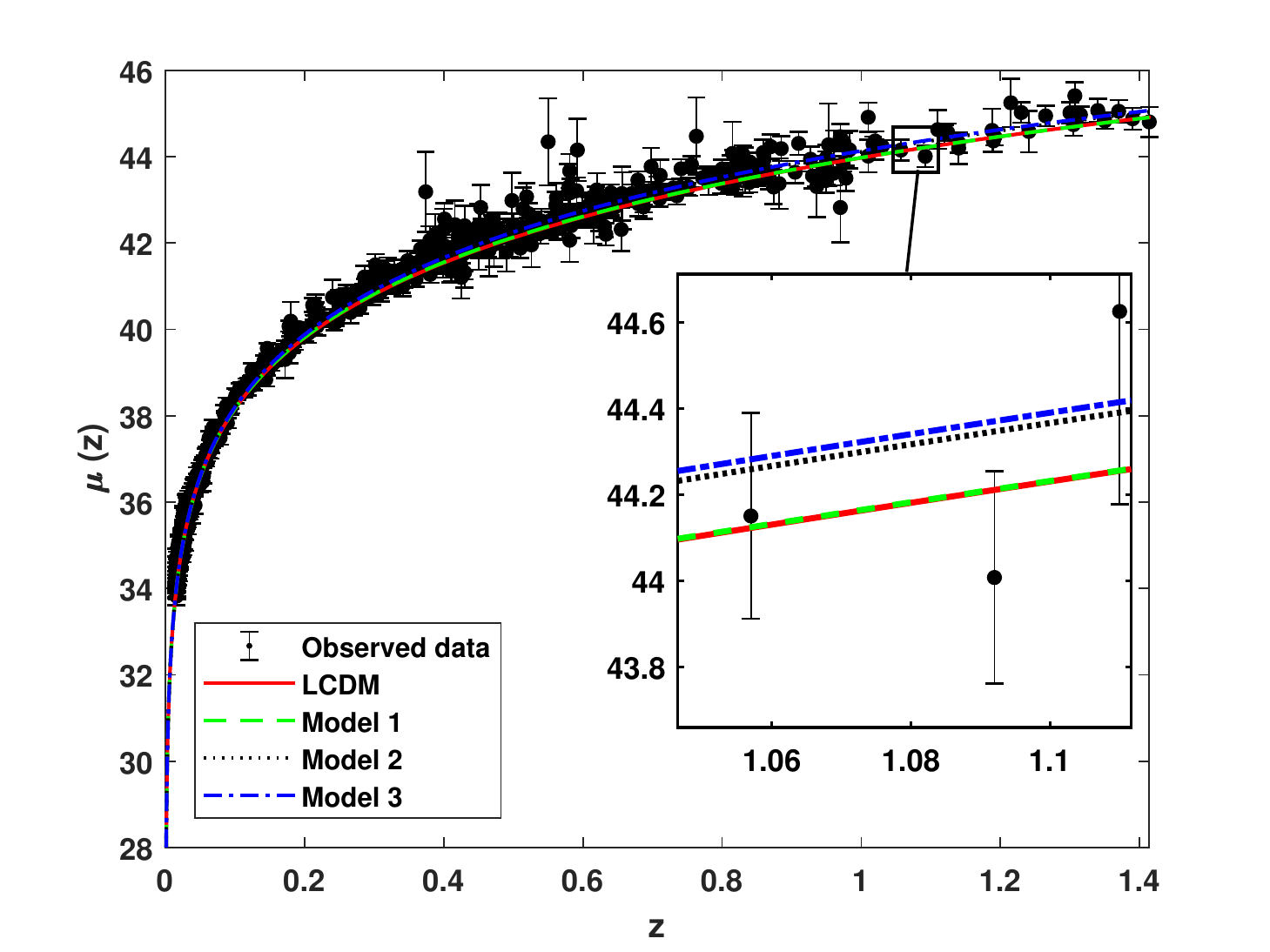}
	\caption{Comparison of the theoretical Hubble parameter 
		(left panel) and the Supernova distance modulus (right panel) computed for different HDE and $\Lambda$CDM models respectively. On top of that we plot the corresponding data. The cosmological parameters are provided in Table (\ref{tab:bestfit}).  
		The style of the different lines are explained in the inner panel of the figure.}
	\label{fig:SNIa}
\end{figure*}
\begin{figure*}
	\centering
	\includegraphics[width=8.5cm]{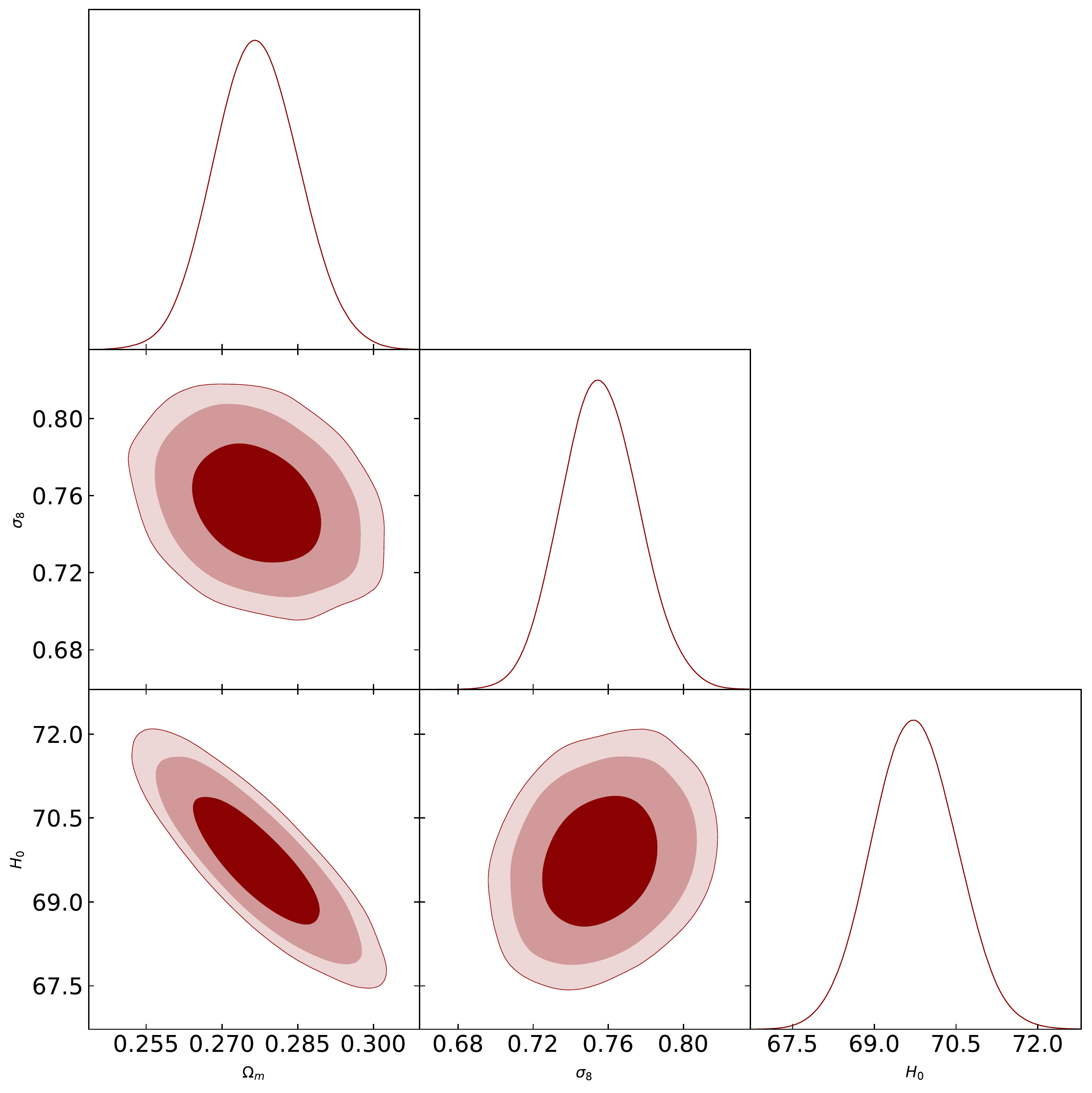}
	\includegraphics[width=8.5cm]{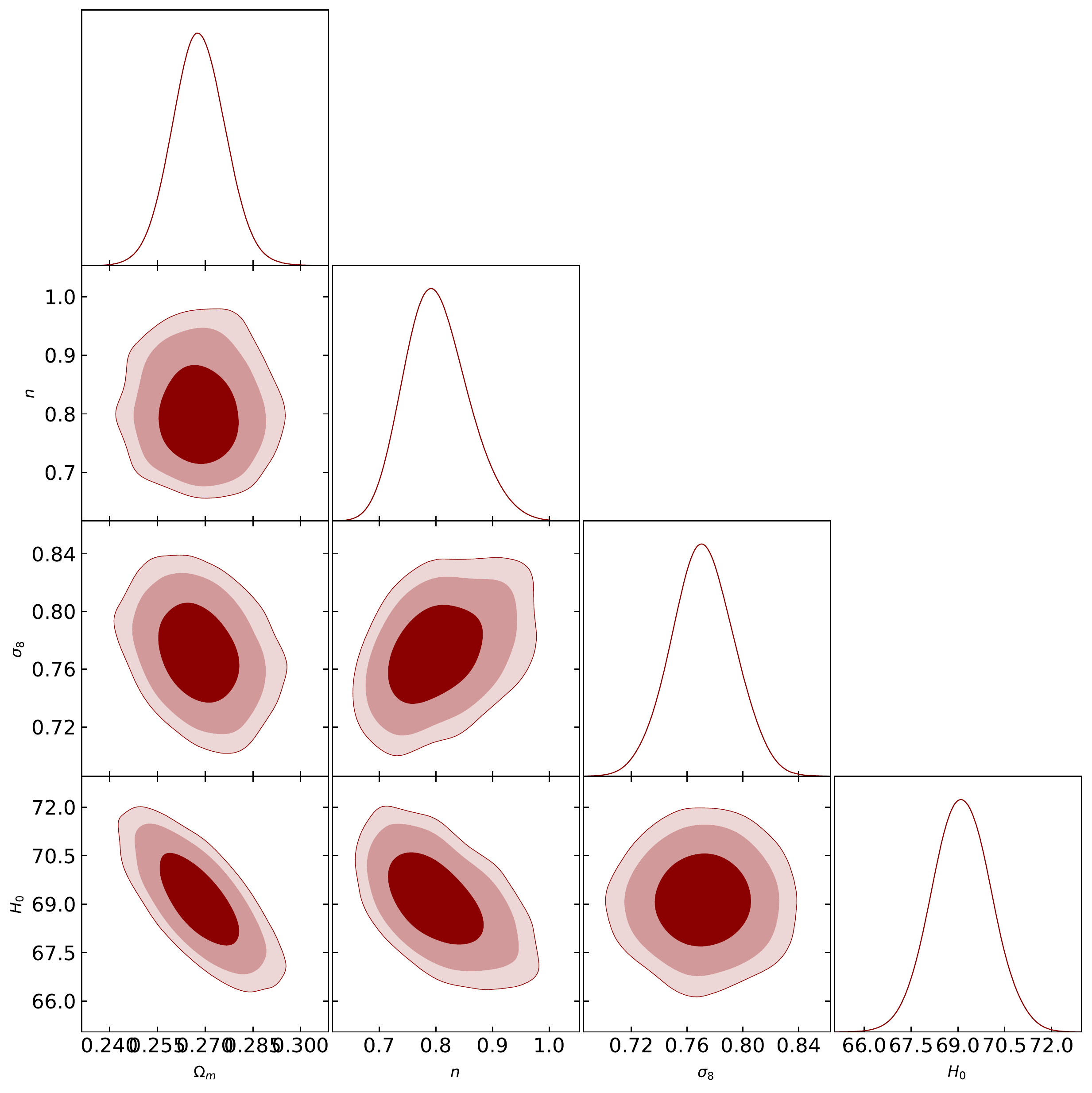}
	\includegraphics[width=8.5cm]{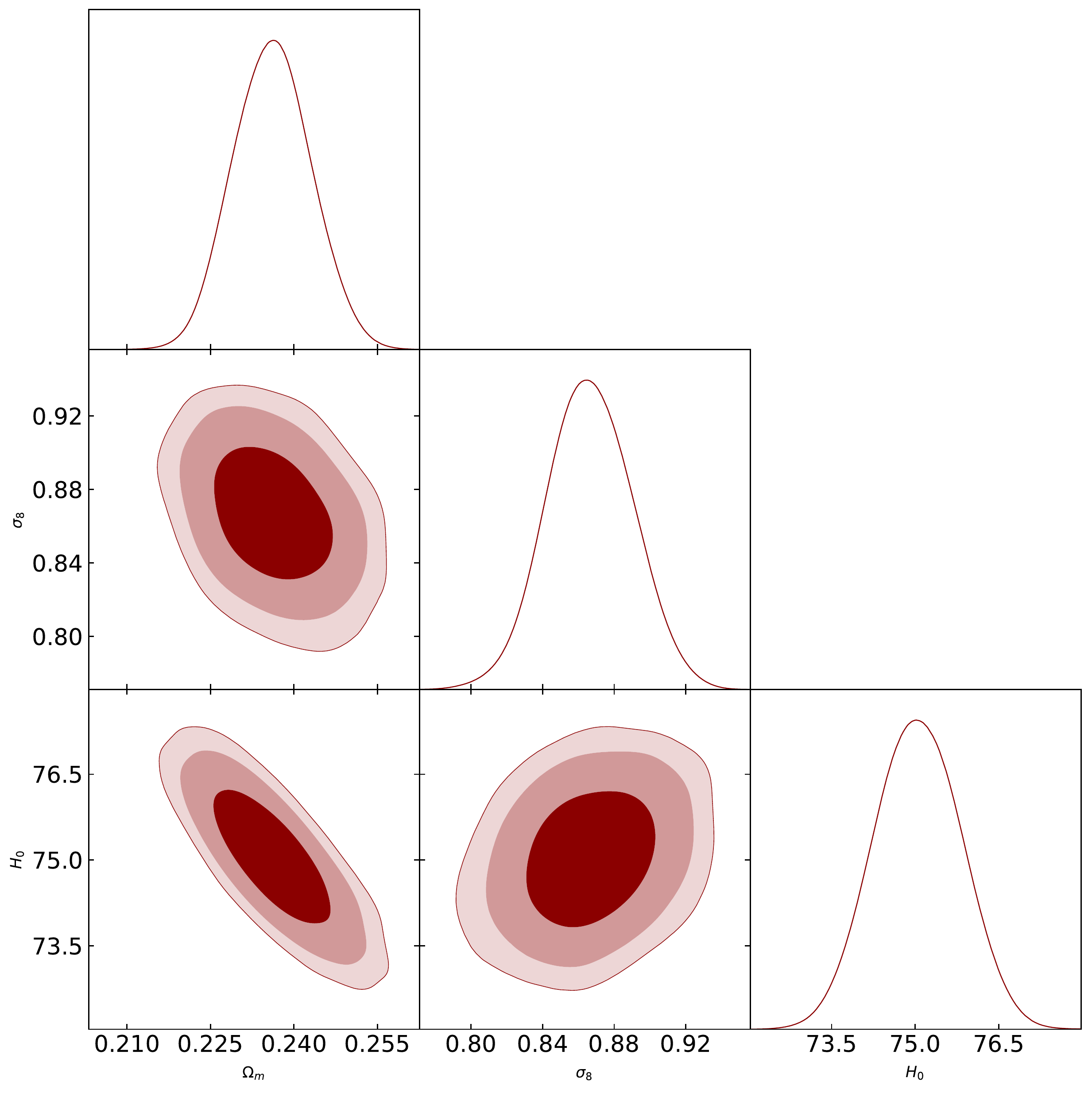}
	\includegraphics[width=8.5cm]{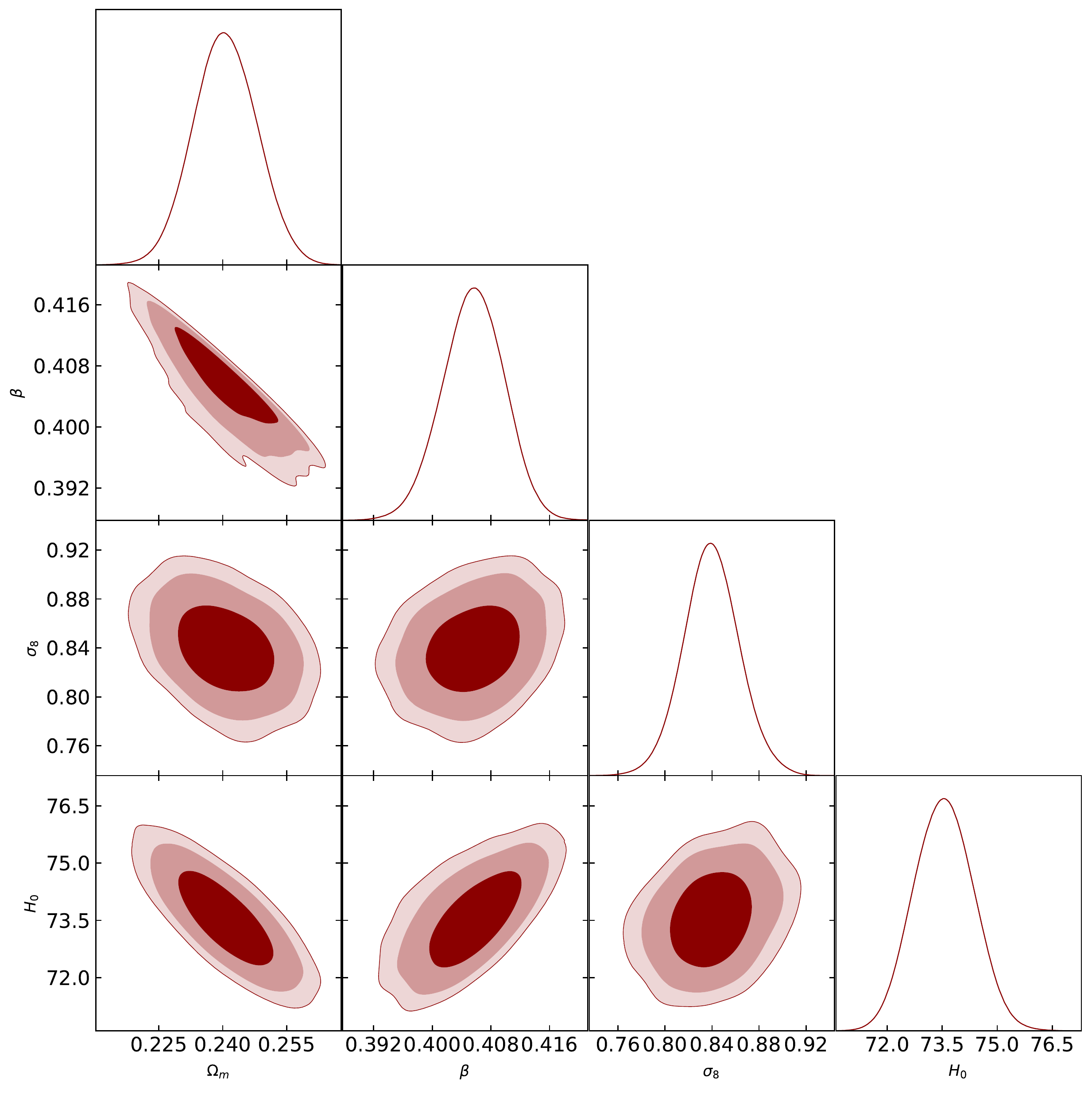}
	\caption{ The $1\sigma$, $2\sigma$ and $3\sigma$ contours 
		for various cosmological parameters. 
		Here we used 
	        the combined background and growth rate data, assuming the HDE is 
		homogeneous (see model 1: up-right, model 2: bottom-left and 
		model 3: bottom right). The concordance $\Lambda$CDM is shown in up-left panel.}
	\label{fig:back+growth1}
\end{figure*}
\begin{figure*}
	\centering
	\includegraphics[width=8.5cm]{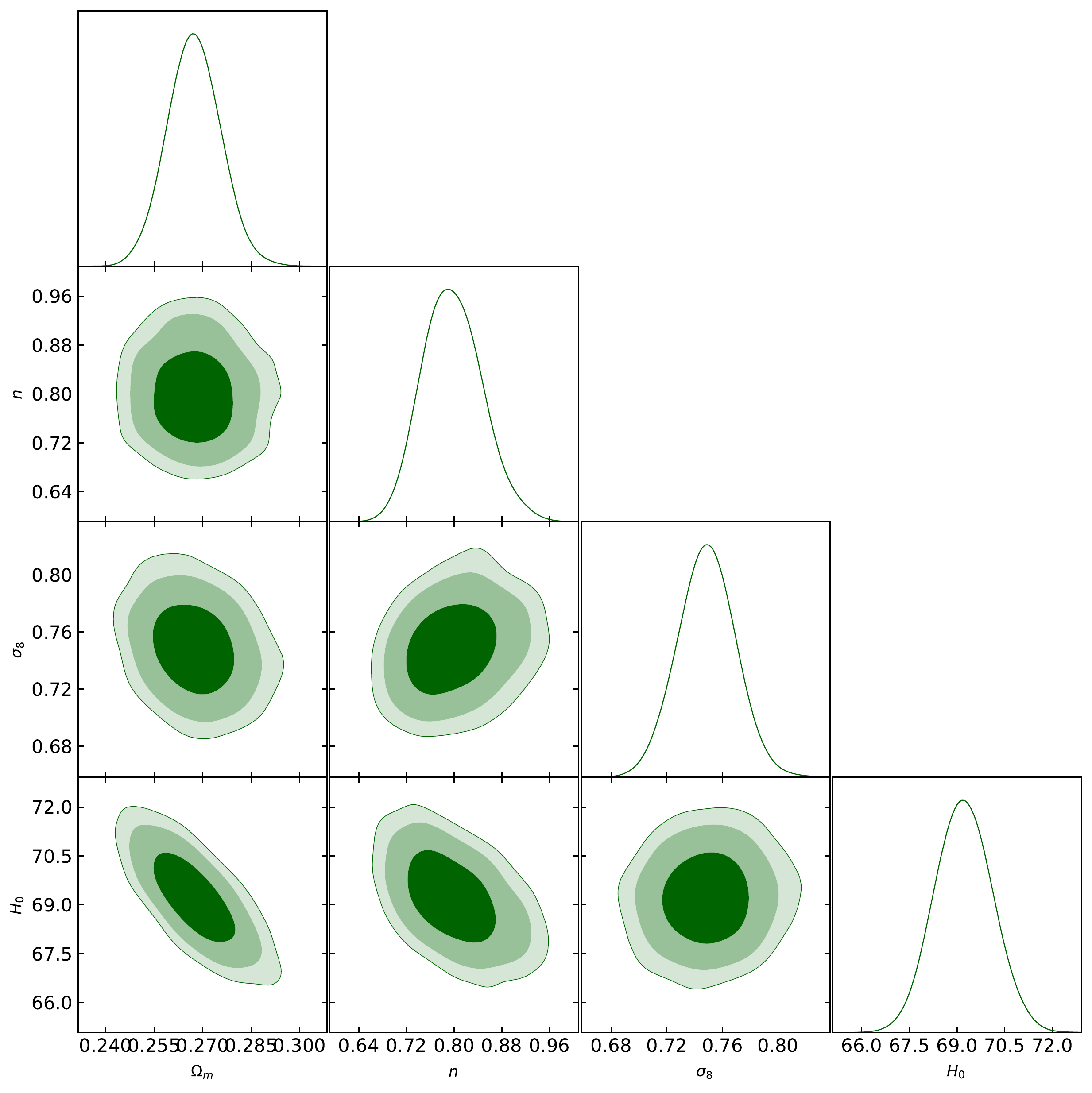}
	\includegraphics[width=8.5cm]{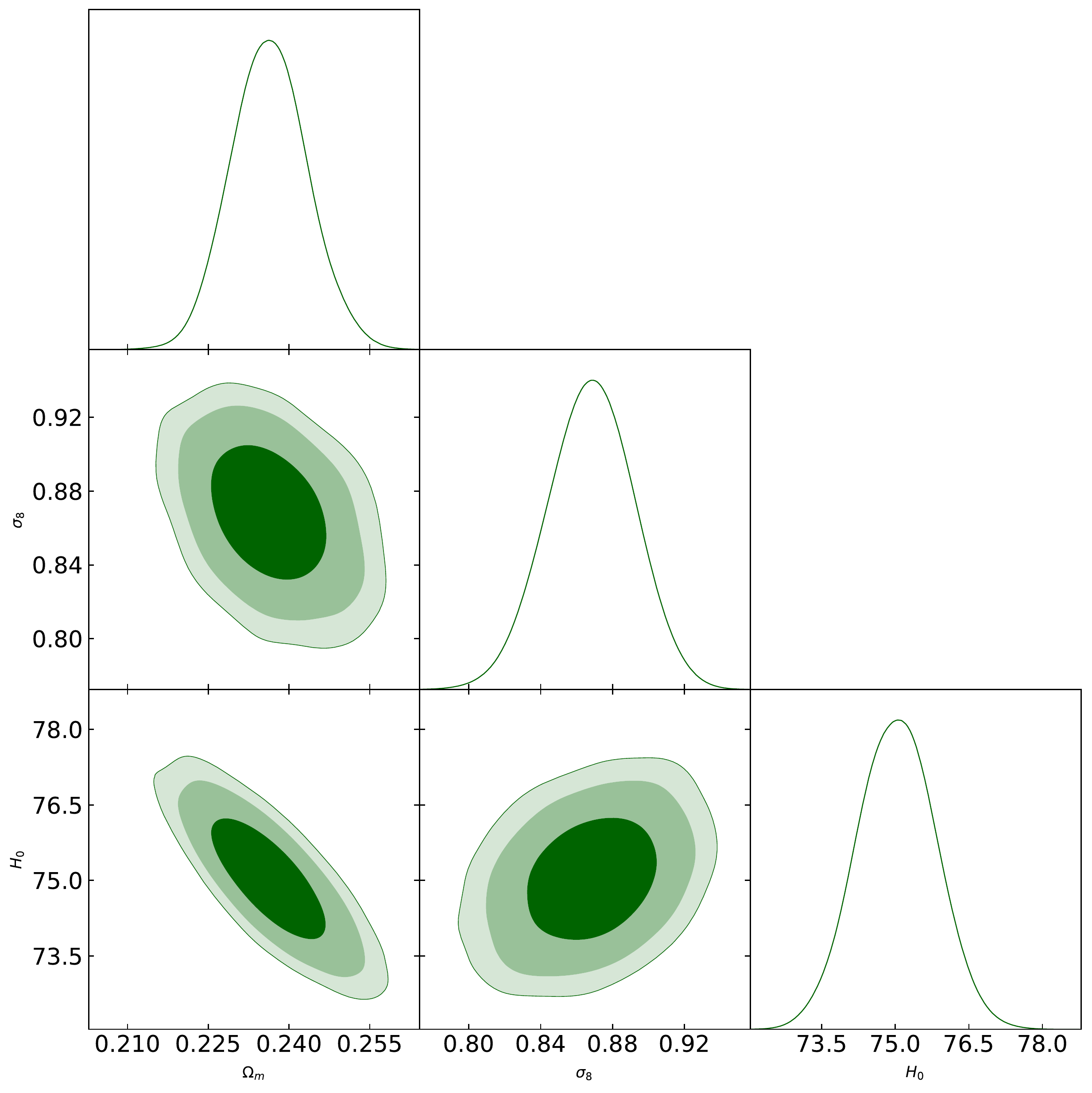}
	\includegraphics[width=8.5cm]{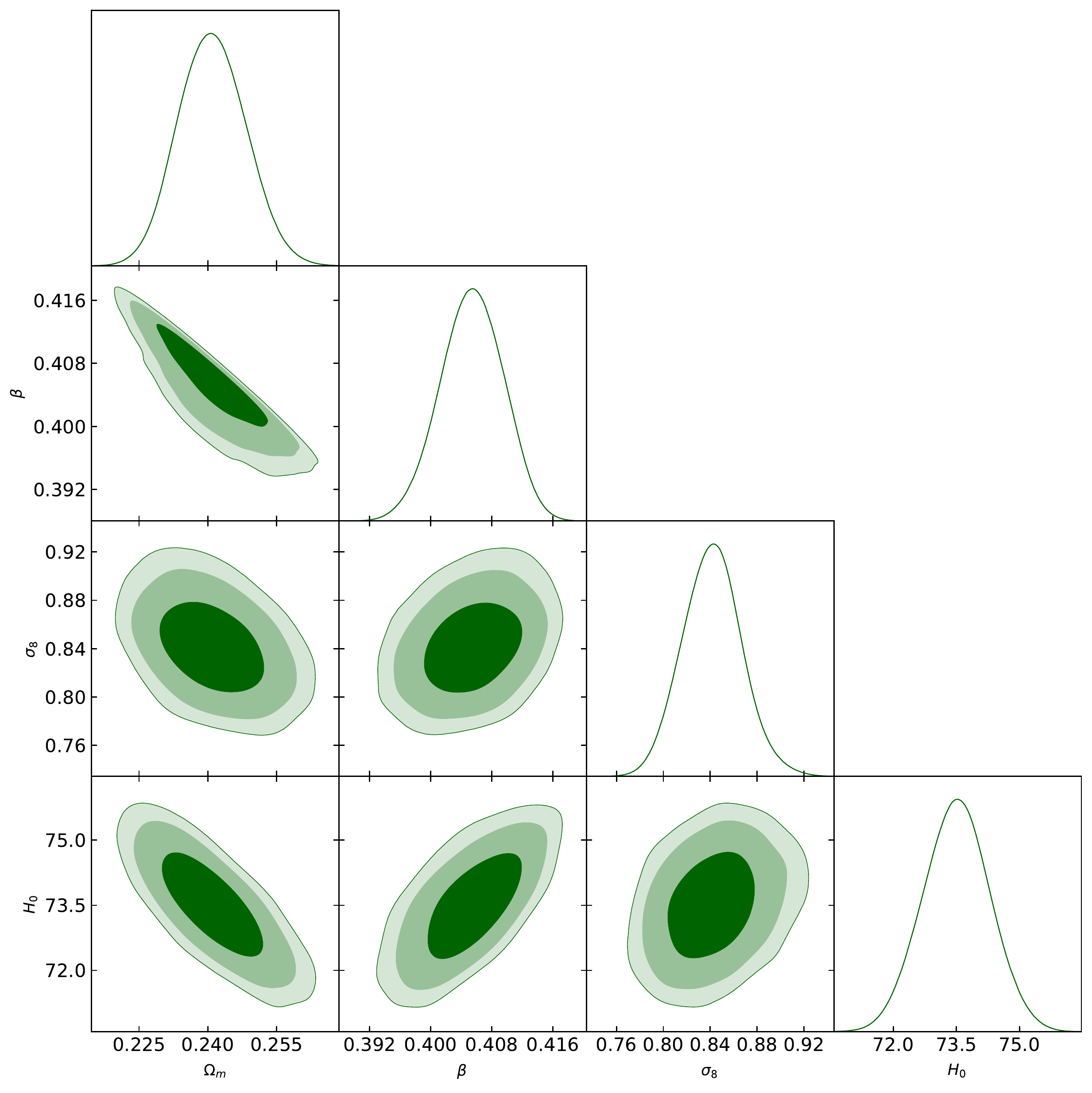}
	\caption{The $1\sigma$, $2\sigma$ and $3\sigma$ contours, 
		considering that the HDE is allowed 
		to cluster (see model 1: up-left panel, model 2: up-right panel and 
		model 3: bottom).}
	\label{fig:back+growth2}
\end{figure*}

\begin{figure*}
	\centering
	\includegraphics[width=5.8cm]{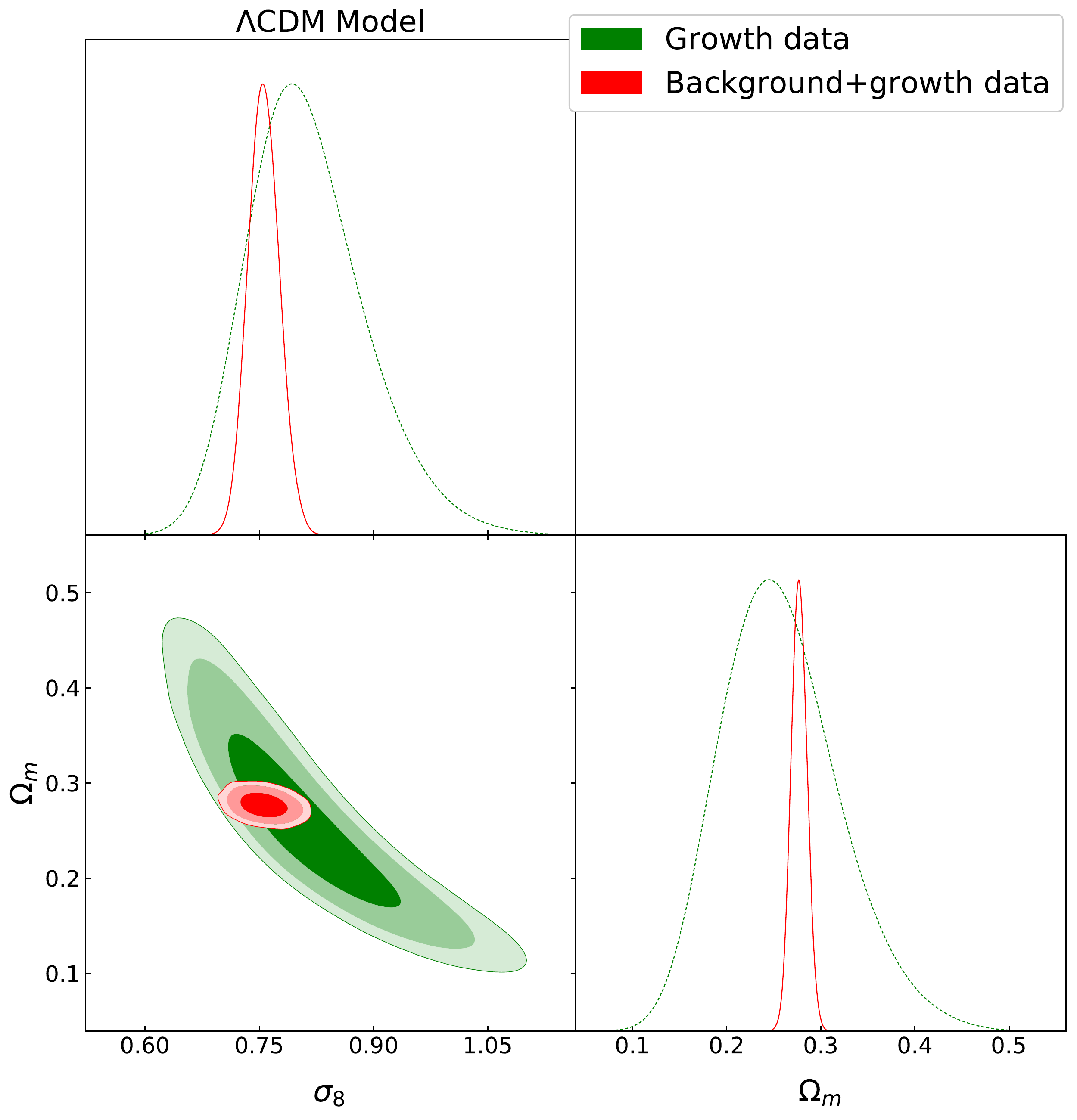}
	\includegraphics[width=5.8cm]{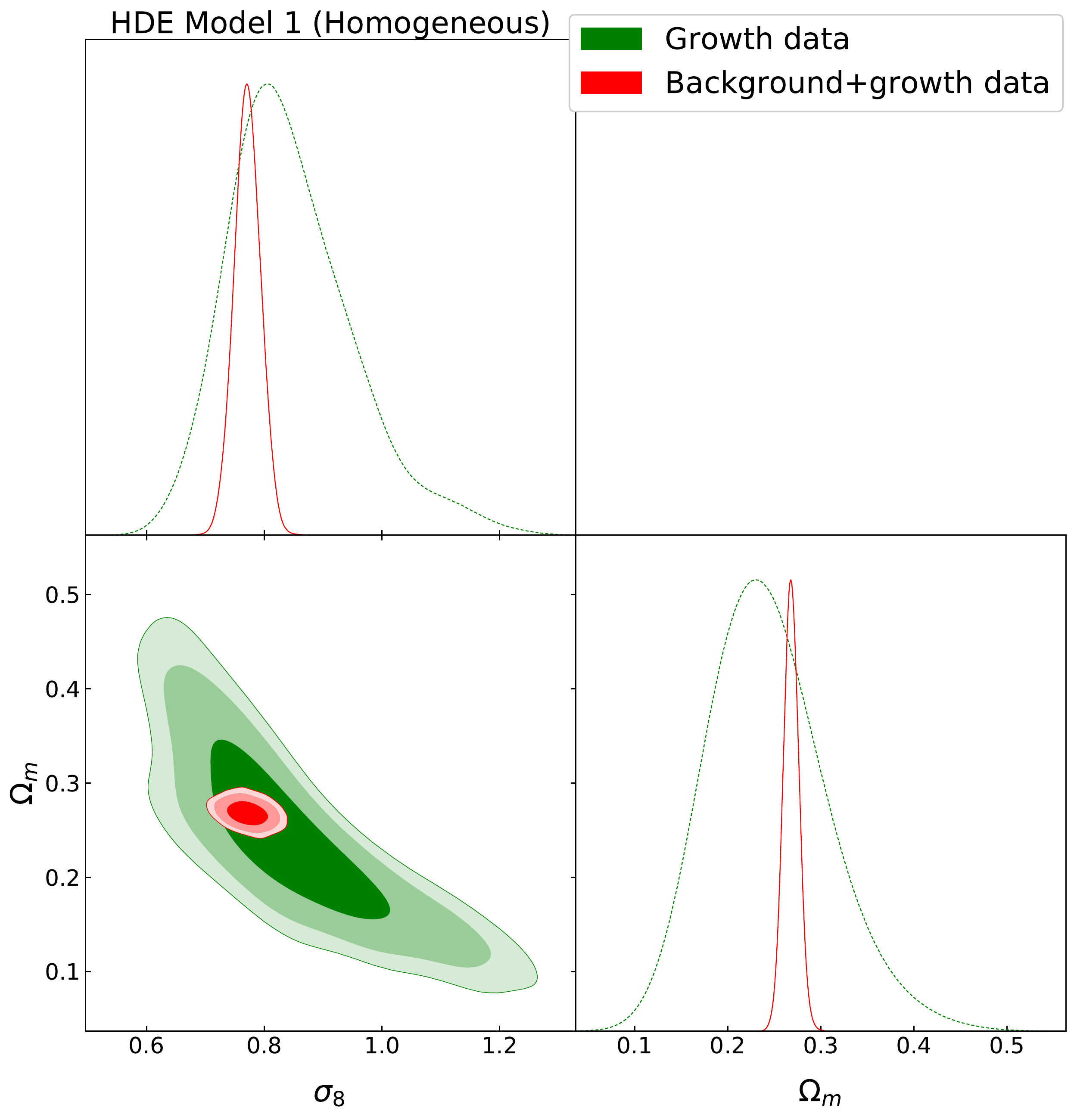}
	\includegraphics[width=5.8cm]{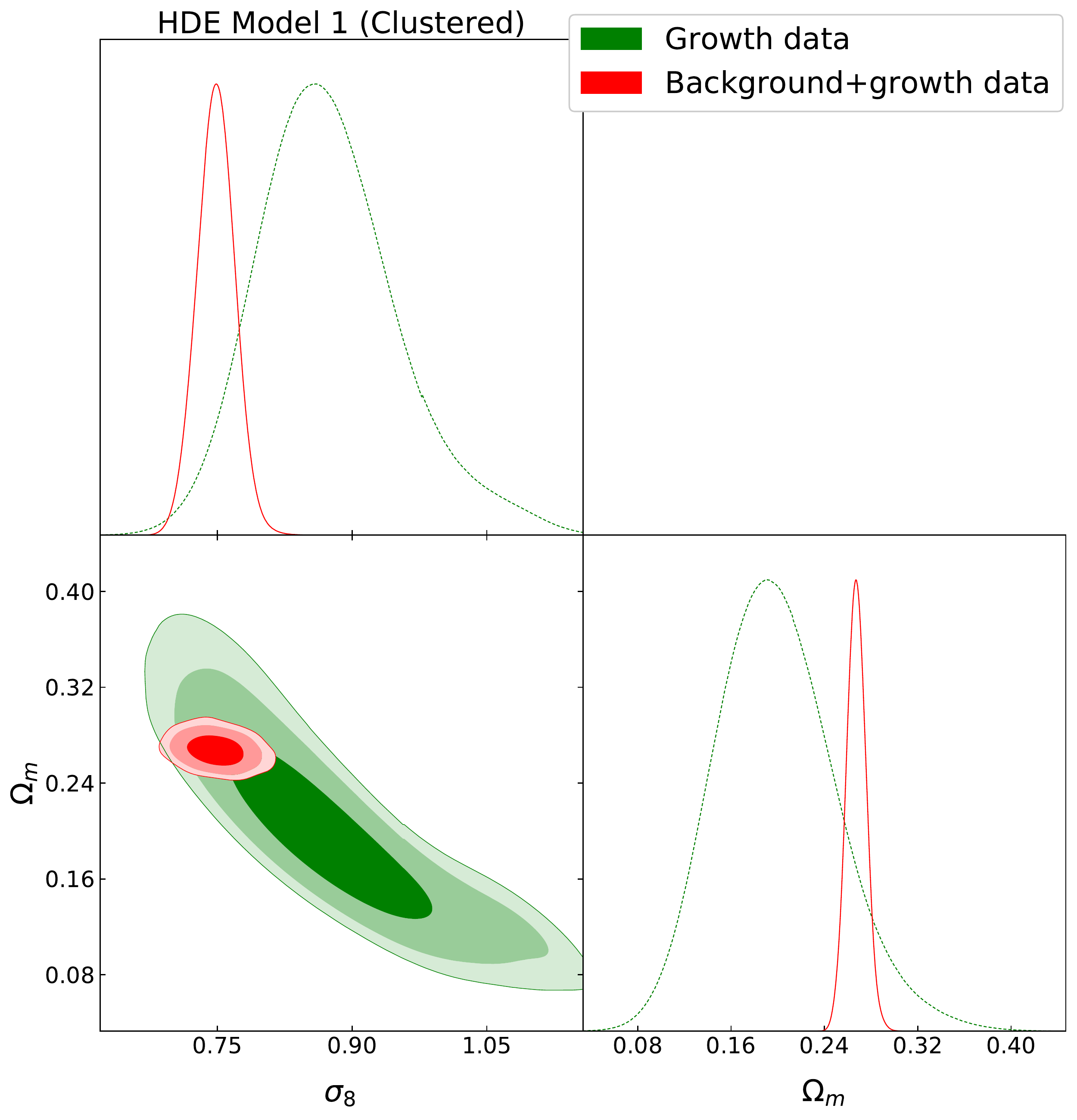}
	\includegraphics[width=5.8cm]{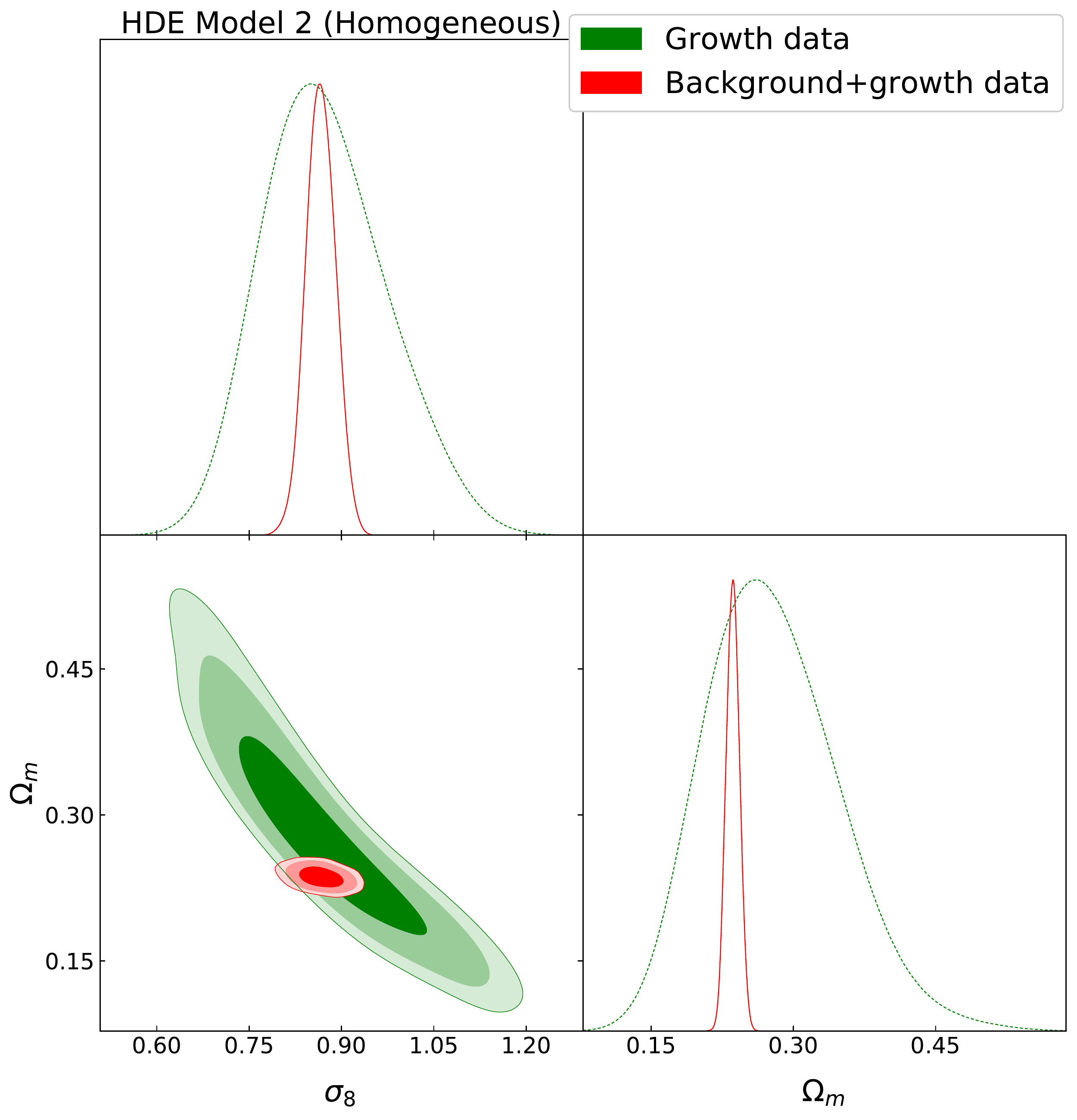}
	\includegraphics[width=5.8cm]{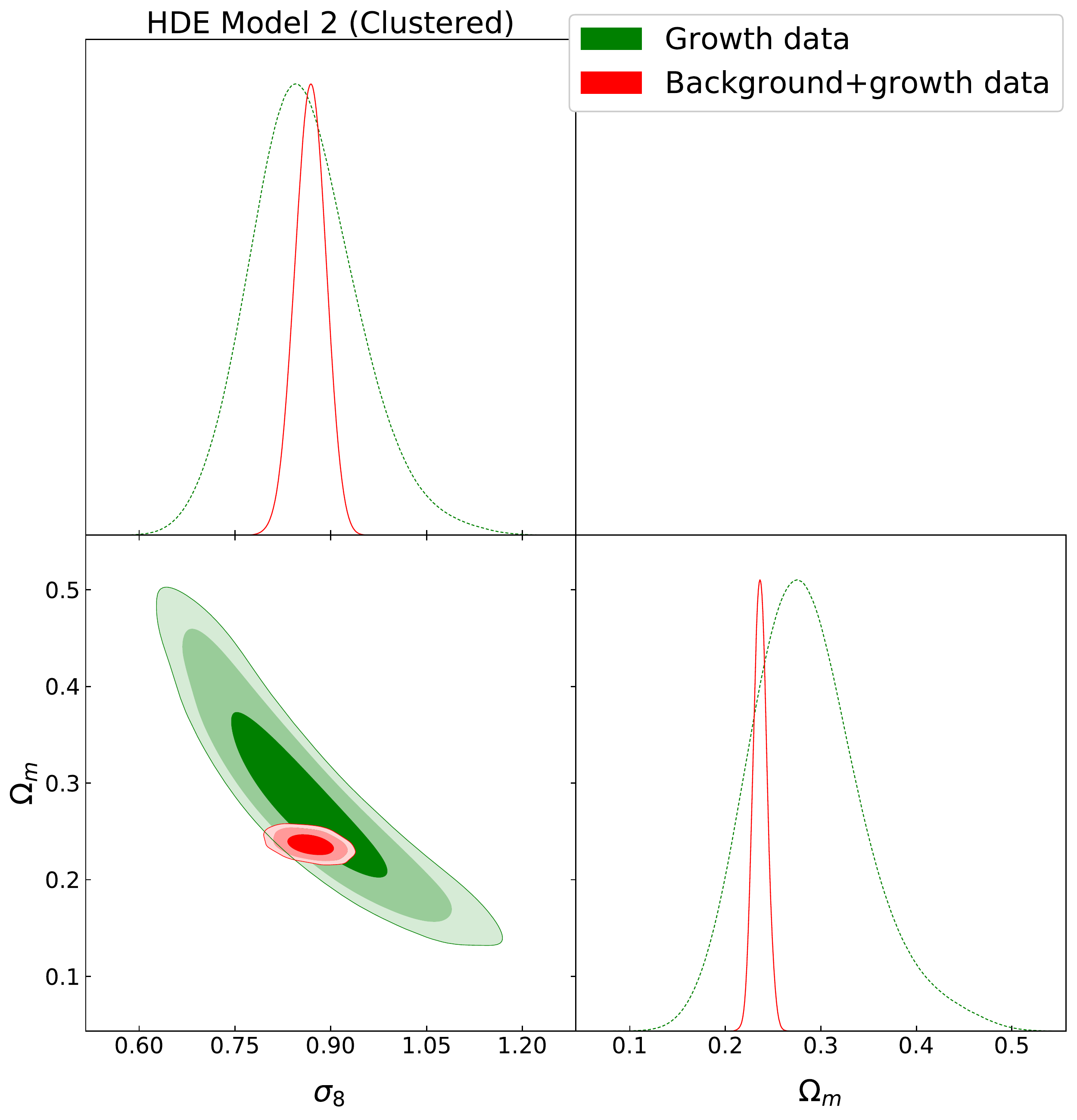}
	\includegraphics[width=5.8cm]{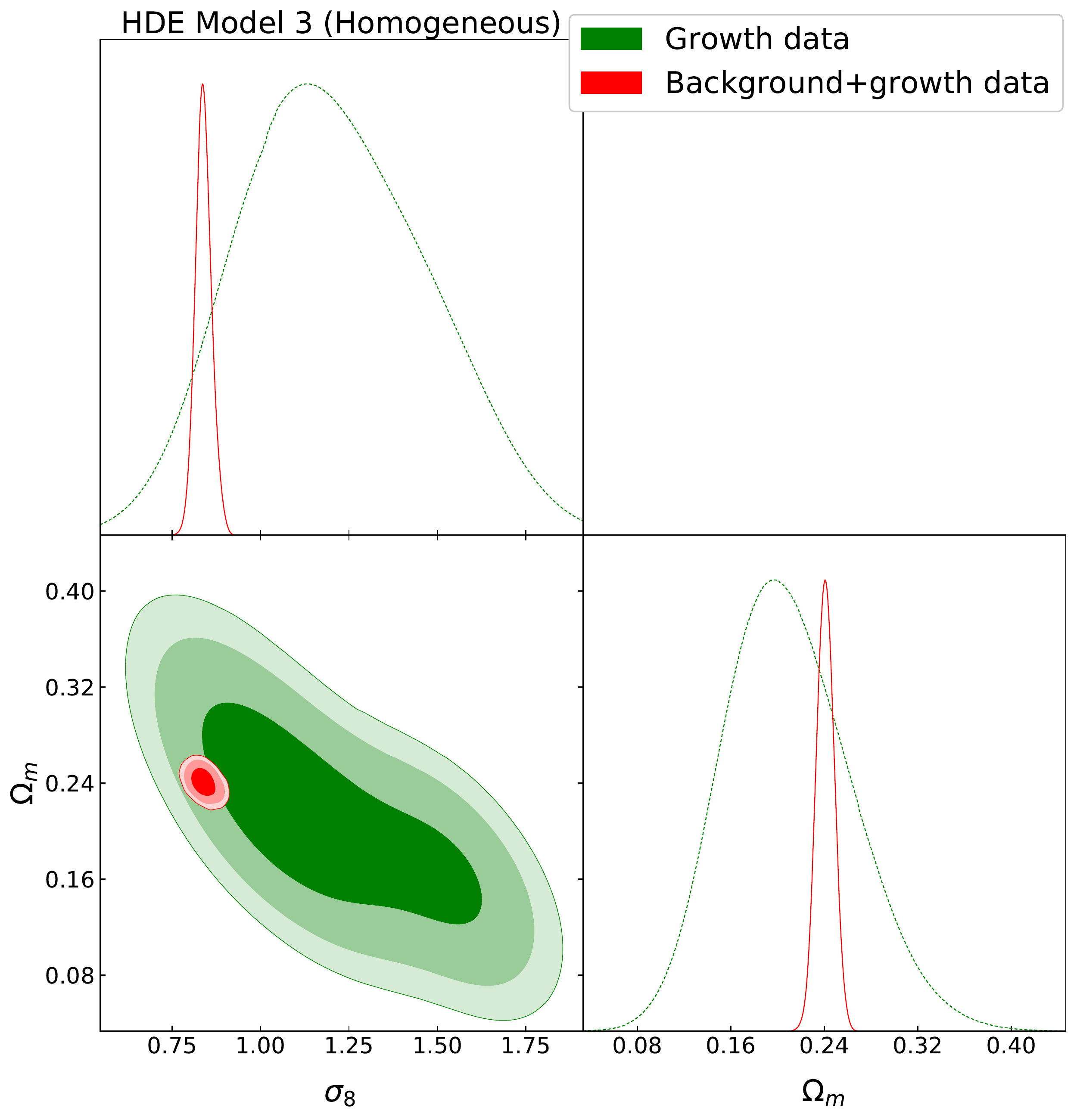}
	\includegraphics[width=5.8cm]{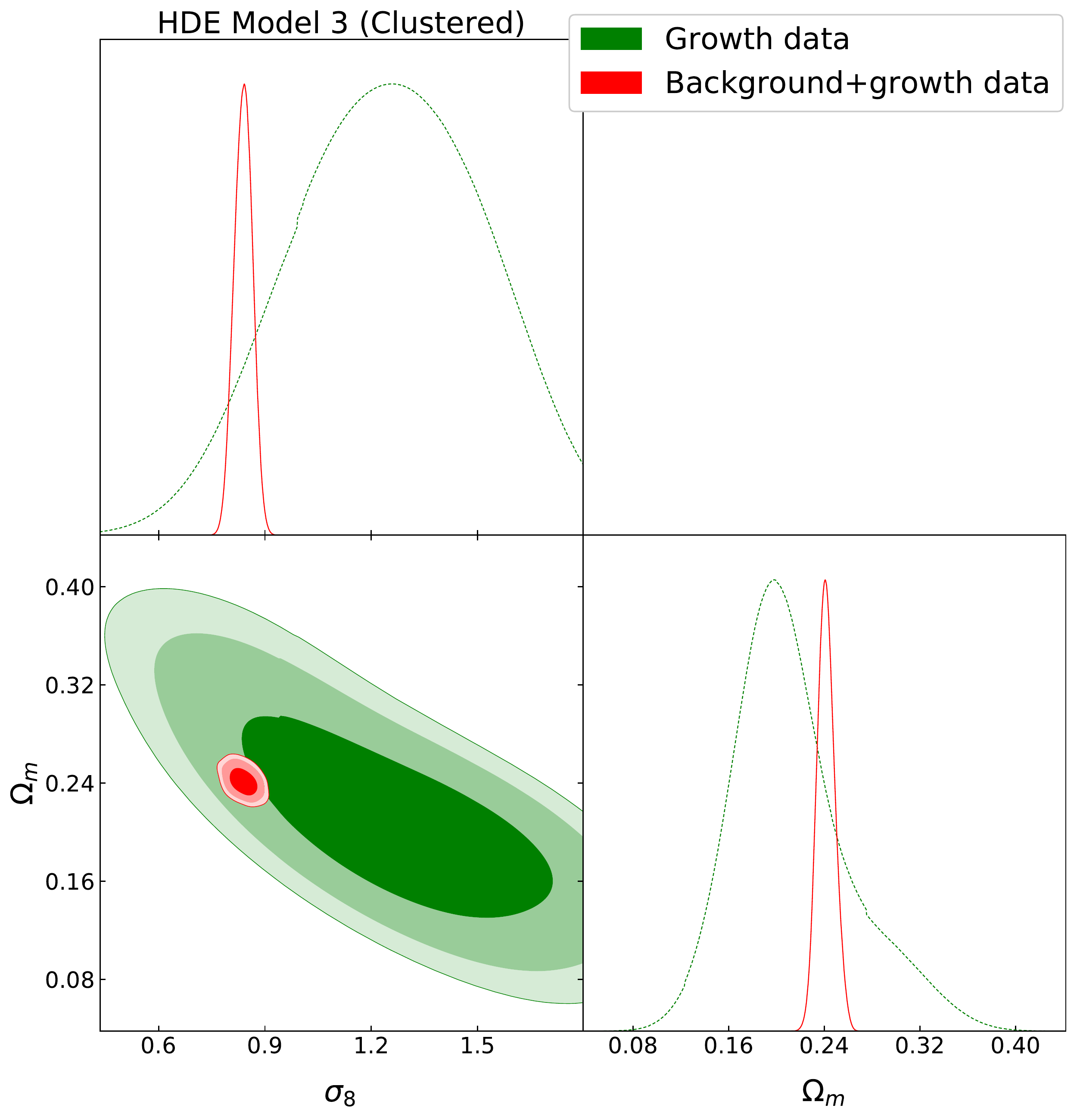}
	\caption{ The $1\sigma$, $2\sigma$ and $3\sigma$ likelihood contours in $\sigma_{8}$- $\Omega_{\rm m}$ 
plane for different HDE models, including the concordance $\Lambda$CDM. The green contours correspond to the growth rate data, while the 
red contours obtained using the combination of expansion and growth data.
}
	\label{fig:back+growth22}
\end{figure*}

\begin{figure} 
	\centering
	\includegraphics[width=8cm]{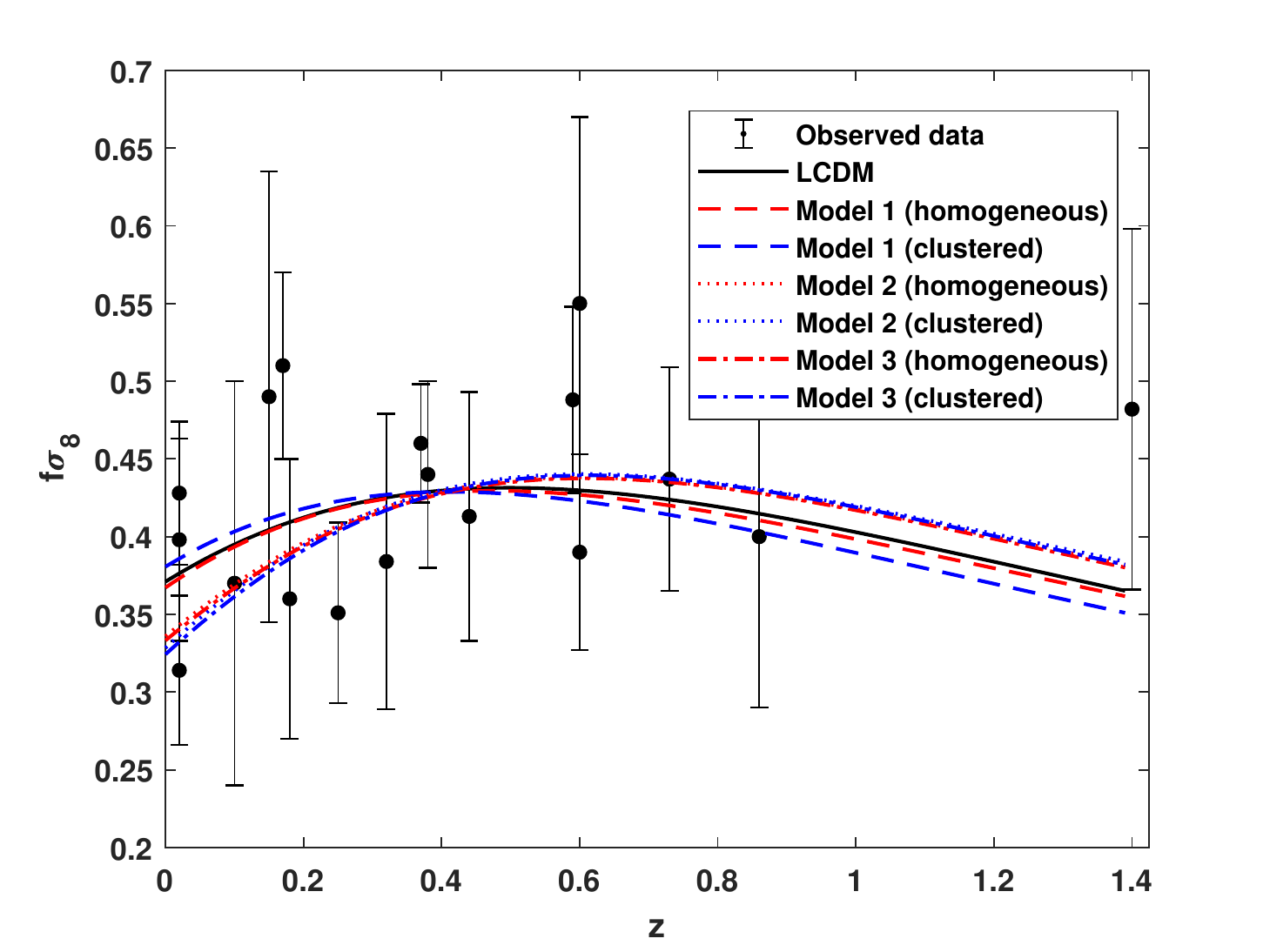}
	\caption{Comparison of the theoretical growth rate for different homogeneous HDE models. On top of that we plot the growth rate data ( see Table 3). For 
		comparison we also include the predictions of the $\Lambda$CDM model. 
		Line styles are explained in the inner plot of the figure. }
	\label{fig:fsigmahom}
\end{figure}

\subsection{Growth of perturbations}
 Here, we briefly review the basic features of the growth 
of linear matter perturbations in DE cosmologies.
We focus our analysis at sub-horizon scales, where the results 
of Pseudo Newtonian dynamics are well consistent with those of General Relativity (GR) paradigm \citep[see][]{Abramo2007}. In this context, 
two different scenarios have been studied in literature \citep{ArmendarizPicon:1999rj,Garriga:1999vw,ArmendarizPicon:2000dh,Erickson:2001bq,Bean:2003fb,Hu:2004yd,Abramo2007,Abramo2008,Ballesteros:2008qk,Abramo:2008ip,Basilakos:2009mz,dePutter:2010vy,Pace2010,Akhoury:2011hr,Sapone:2012nh,Pace2012,Batista:2013oca,Dossett:2013npa,Batista:2014uoa,Basse:2013zua,Pace2014a,Pace:2013pea,Pace:2014taa,Malekjani:2015pza,Naderi2015,Mehrabi:2014ema,Mehrabi:2015hva,Mehrabi:2015kta,Nazari-Pooya:2016bra,Malekjani:2016edh}. 
In the first scenario the DE component is homogeneous  
and only the corresponding non-relativistic matter is allowed to clump, while
in the second scenario the whole system clusters (both matter and DE). 
For both treatments, we refer the reader to follow our previous articles  
\citep{Mehrabi:2015kta,Malekjani:2016edh} in which we have provided 
the basic differential equations which describe the 
situation at the perturbation level. 
Concerning the initial conditions, we use those 
provided by \cite{Batista:2013oca}
\citep[see also][]{Mehrabi:2015kta,Malekjani:2016edh}.
Here we study the growth of matter perturbations from 
the epoch of matter-radiation equality to the present time. 

In the case of homogeneous DE models, DE affects the 
growth of matter perturbations via the Hubble parameter,  
while for clustered DE models, DE affects 
the growth of matter fluctuations through: (i) the modification of the 
Hubble rate and (ii) the direct influence of DE perturbations on 
the matter perturbations. 
Notice, that the DE fluctuations can grow in a similar way to matter perturbations. 
Of course, due to the impact of negative pressure the amplitude of DE perturbations is much smaller 
with respect to that of matter perturbations. 
Moreover, the influence of DE perturbations on the growth 
of matter fluctuations depends on the EoS parameter of DE. 
Indeed in the case of DE models with quintessence 
like EoS $-1<w_{\rm de}<-1/3$, DE perturbations causes the 
decrement of the amplitude of matter fluctuations \citep{Abramo2007}. 
On the other hand for phantom DE models ($w_{\rm de}<-1$), DE 
perturbations enhance the process of matter fluctuations growth \citep{Abramo2007}. In the following section, we consider both clustered and 
homogeneous HDE models and we compare 
the predicted growth rate of matter perturbations against the data.

\begin{table}\label{tab:fff}
	\caption{ The $H(z)$ data. Notice that the values of $H(z)$ are in units [km/s/Mpc]}
	\begin{tabular}{c  c  c }
		\hline \hline
		z & H(z) & References \\
		\hline
		$0.07$ & $69.0^{+19.6}_{-19.6}$ & \cite{Zhang:2012mp}\\
		
		$0.09$ & $69.0^{+12.0}_{-12.0}$ & \cite{Jimenez:2003iv}\\
		
		$0.12$ & $69.0^{+12.0}_{-12.0}$  &  \cite{Zhang:2012mp}\\	
		
		$0.17$ & $83.0^{+8.0}_{-8.0}$  &  \cite{Simon:2004tf}\\
		
		$0.1791$ & $75.0^{+4.0}_{-4.0}$  &  \cite{Moresco:2012jh}\\
		
		$0.1993$ & $75.0^{+5.0}_{-5.0}$  &  \cite{Moresco:2012jh}\\
		
		$0.2$ & $72.9^{+29.6}_{-29.6}$  &  \cite{Zhang:2012mp}\\
		
		$0.27$ & $77.0^{+14}_{-14}$  &  \cite{Simon:2004tf}\\
		
		$0.28$ & $88.8^{+36.6}_{-36.6}$  &  \cite{Zhang:2012mp}\\
		
		$0.3519$ & $83.0^{+14.0}_{-14.0}$  &  \cite{Moresco:2012jh}\\
		
		$0.3802$ & $83.0^{+13.5}_{-13.5}$  &  \cite{Moresco:2016mzx}\\
		
		$0.4$ & $95.0^{+17.0}_{-17.0}$  &  \cite{Simon:2004tf}\\
		
		$0.4004$ & $77.0^{+10.2}_{-10.2}$  &  \cite{Moresco:2016mzx}\\
		
		$0.4247$ & $87.1^{+11.2}_{-11.2}$  &  \cite{Moresco:2016mzx}\\
		
		$0.4497$ & $92.8^{+12.9}_{-12.9}$  &  \cite{Moresco:2016mzx}\\
		
		$0.4783$ & $80.9^{+9.0}_{-9.0}$  &  \cite{Moresco:2016mzx}\\
		
		$0.48$ & $97.0^{+62.0}_{-62.0}$  &  \cite{Stern:2009ep}\\
		
		$0.5929$ & $104.0^{+13.0}_{-13.0}$  &  \cite{Moresco:2012jh}\\
				
		$0.6797$ & $92.0^{+8.0}_{-8.0}$  &  \cite{Moresco:2012jh}\\
		
		$0.7812$ & $105.0^{+12.0}_{-12.0}$  &  \cite{Moresco:2012jh}\\
		
		$0.8754$ & $125.0^{+17.0}_{-17.0}$  &  \cite{Moresco:2012jh}\\
		
		$0.88$ & $90.0^{+40.0}_{-40.0}$  &  \cite{Stern:2009ep}\\
		
		$0.90$ & $117.0^{+23.0}_{-23.0}$  &  \cite{Simon:2004tf}\\
		
		$1.037$ & $154.0^{+20.0}_{-20.0}$  &  \cite{Moresco:2012jh}\\
		
		$1.3$ & $168.0^{+17.0}_{-17.0}$  &  \cite{Moresco:2012jh}\\
		
		$1.363$ & $160.0^{+33.0}_{-33.0}$  &  \cite{Moresco:2015cya}\\
		
		$1.43$ & $177.0^{+18.0}_{-18.0}$  &  \cite{Simon:2004tf}\\
		
		$1.53$ & $140.0^{+14.0}_{-14.0}$  &  \cite{Simon:2004tf}\\
		
		$1.75$ & $202.0^{+40.0}_{-40.0}$  &  \cite{Simon:2004tf}\\
		
		$1.965$ & $186.0^{+50.4}_{-50.4}$  &  \cite{Moresco:2015cya}\\
		\hline\hline
	\end{tabular}
\end{table}

\begin{table}\label{tab:ttt}
	\caption{ The growth rate $f(z)\sigma_8(z)$ data. }
	\begin{tabular}{c  c  c }
		\hline \hline
		$z$ & $f(z)\sigma_8(z)$ & References \\
		\hline
		$0.02$ & $0.428^{+0.0465}_{-0.0465}$ & \cite{Huterer:2016uyq}\\
		
		$0.02$ & $0.398^{+0.065}_{-0.065}$ & \cite{Hudson:2012gt},\cite{Turnbull:2011ty}\\
		
		$0.02$ & $0.314^{+0.048}_{-0.048}$ & \cite{Hudson:2012gt},\cite{Davis:2010sw}\\
		
		$0.10$ & $0.370^{+0.130}_{-0.130}$ & \cite{Feix:2015dla}\\
		
		$0.15$ & $0.490^{+0.145}_{-0.145}$ & \cite{Howlett:2014opa}\\
		
		$0.17$ & $0.510^{+0.060}_{-0.060}$ & \cite{Song:2008qt}\\
		
		$0.18$ & $0.360^{+0.090}_{-0.090}$ & \cite{Blake:2013nif}\\
		
		$0.38$ & $0.440^{+0.060}_{-0.060}$ & \cite{Blake:2013nif}\\
		
		$0.25$ & $0.3512^{+0.0583}_{-0.0583}$ & \cite{Samushia:2011cs}\\
		
		$0.37$ & $0.4602^{+0.0378}_{-0.0378}$ & \cite{Samushia:2011cs}\\
		
		$0.32$ & $0.384^{+0.095}_{-0.095}$ & \cite{Sanchez:2013tga}\\
		
		$0.59$ & $0.488^{+0.060}_{-0.060}$ & \cite{Chuang:2013wga}\\
		
		$0.44$ & $0.413^{+0.080}_{-0.080}$ & \cite{Blake:2012pj}\\
		
		$0.60$ & $0.390^{+0.063}_{-0.063}$ & \cite{Blake:2012pj}\\
		
		$0.73$ & $0.437^{+0.072}_{-0.072}$ & \cite{Blake:2012pj}\\
		
		$0.60$ & $0.550^{+0.120}_{-0.120}$ & \cite{Pezzotta:2016gbo}\\
		
		$0.86$ & $0.400^{+0.110}_{-0.110}$ & \cite{Pezzotta:2016gbo}\\
		
		$1.40$ & $0.482^{+0.116}_{-0.116}$ & \cite{Okumura:2015lvp}\\
		\hline \hline
	\end{tabular}
\end{table}

\begin{table*}
	\centering
	\caption{Results of statistical likelihood analysis using different sets of background data for various HDE models and standard $\Lambda$CDM universe.
	}
	\begin{tabular}{c  c  c c c c}
		\hline \hline
		Model  & Model (1)& Model (2) & Model (3) &$\Lambda$CDM\\
		\hline
		$k$ & 4& 3& 4& 3\\
		\hline
		$\chi^2_{\rm min}$(total) & 591.28 & 728.52 & 657.56 & 587.64& \\
		\hline
		$\chi^2_{\rm best}$(SNIa) & 562.43 & 600.53 & 609.09 &  562.23& \\
		\hline
		$\chi^2_{\rm best}$(Hubble) & 22.04 & 48.17 & 28.85 &  20.63& \\
		\hline
		$\chi^2_{\rm best}$(BBN) & 0.18 & 3.84 & 0.68 &  0.02& \\
		\hline
		$\chi^2_{\rm best}$(CMB: WMAP data) & 2.25 &50.98 & 6.66 &  0.59& \\
		\hline
		$\chi^2_{\rm best}$(BAO) & 4.37 & 25.00 & 12.29 &  4.17 \\
		\hline
		AIC  & 599.28 & 734.52 & 665.56 & 593.64 & \\
		\hline
		BIC  & 616.04 & 747.84 & 683.32 & 606.96 & \\
		\hline \hline
	\end{tabular}\label{tab:best}
\end{table*}

\begin{table*}
	\centering
	\caption{Best-fit parameters for the various HDE models using the cosmological data at background level.}
	\begin{tabular}{c  c  c c c c}
		\hline \hline
		Model  & Model(1)& Model(2)& Model(3)& $\Lambda$CDM\\
		\hline
		$\Omega_{\rm m}$ & $0.2677_{-0.0082,-0.016,-0.021}^{+0.0082,+0.016,+0.022}$& $\Omega_m = 0.2344_{-0.0070,-0.013,-0.017}^{+0.0070,+0.014,+0.019}$ & $0.2438_{-0.0075,-0.014,-0.019}^{+0.0075,+0.015,+0.020} $ & $0.2767_{-0.0083,-0.016,-0.021}^{+0.0083,+0.017,+0.022}$ \\
		\hline
		$ H_0 $& $69.29_{-0.92,-1.80,-2.40}^{+0.92,+1.80,+2.30}$ &$75.20_{-0.79,-1.50,-2.10}^{+0.79,+1.50,+2.00}$ &
		$71.66_{-0.90,-1.70,-2.30}^{+0.90,+1.80,2.40}$ &  $69.74_{-0.77,+1.50,+1.90}^{+0.77,-1.50,-1.90}$&\\
		\hline
		$ n $ & $ 0.785^{+0.042,+0.10,+0.15}_{-0.056,-0.094,-0.11} $&-- & --&--& \\
		\hline
		$ \beta$ &--&--& $0.4369_{-0.0090,-0.016,-0.022}^{+0.0090,+0.017,+0.023}$&--&\\
		\hline
		$ w_{\rm de}(z=0) $ & -1.10 & -1.29 & -1.32 & -1.00\\
		\hline
		$ \Omega_{\rm de}(z=0) $ & 0.71372 & 0.77314 & 0.75447 & 0.72627 \\
		\hline \hline
	\end{tabular}\label{tab:bestfit}
\end{table*}

\begin{table*}
	\centering
	\caption{ Numerical results for different homogeneous HDE models (part A) and clustered HDE models (part B)  using the growth rate data. The results for concordance $\Lambda$CDM universe are shown for comparison.
	}
	\begin{tabular}{c  c  c c c c}
		\hline \hline
		Part (A)  & Model 1 (homogeneous) & Model 2 (homogeneous) & Model 3 (homogeneous) & $\Lambda$CDM\\
		\hline
		$\chi^2_{\rm min} (gr)$ & 11.2 & 11.9 & 11.1 & 11.5 \\
		\hline 
		AIC (BIC) & 19.2 (19.6) & 17.9 (18.8) & 19.1 (19.5) & 17.5 (18.4)\\
		\hline
		$\Omega_{\rm m}$ & $0.242_{-0.070,-0.12,-0.14}^{+0.055,+0.13,+0.20}$&  $0.277_{-0.077,-0.013,-0.016}^{+0.061,+0.14,+0.20}$  & $0.216_{-0.083,-0.12,-0.14}^{+0.052,+0.14,+0.22} $  & $0.257_{-0.07,-0.12,-0.13}^{+0.052,+0.13,+0.18}$ \\
		\hline
		$ \sigma_8 $& $0.844_{-0.12,-0.20,-0.22}^{+0.080,+0.23,+0.35} $&$0.872_{-0.11,-0.18,-0.20}^{+0.089,+0.21,+0.26}$ & $1.27_{-0.48,-0.54,-0.57}^{+0.24,+0.78,+0.87}$ &  $0.813_{-0.087,-0.14,-0.16}^{+0.061,+0.16,+0.24}$&\\
			\hline \hline
			Part (B)  & Model 1 (clustered) & Model 2 (clustered) & Model 3 (clustered) \\
			\hline
			$\chi^2_{\rm min} (gr)$ & 11.2 & 12.0 & 11.1   \\
			\hline 
			AIC (BIC) & 19.2(19.6) & 18.0 (18.9) & 19.1 (19.6) \\
			\hline
			$\Omega_{\rm m}$ & $0.198^{+0.046 , +0.10, +0.15}_{-0.054 , -0.096 , -0.12}$& $0.285^{+0.048 , +0.12 , +0.18}_{-0.065 , -0.11 , -0.13}$  & $0.212^{+0.034 , +0.12 , +0.15}_{-0.056 -0.84 , -0.094} $   \\
			\hline
			$ \sigma_8 $& $0.873^{+0.064 , +0.17 , +0.22}_{-0.085 , -0.15 , -0.17} $&$0.858^{+0.074 , +0.18 , +0.26}_{-0.089 , -0.16 , -0.19}$ & $1.25^{+0.32 , +0.52 , +0.64}_{-0.28 , -0.51 , -0.63}$ \\ 
			\hline \hline
	\end{tabular}\label{tab:bestfsigma1}
\end{table*}

\begin{table*}
	\centering
	\caption{Results of statistical likelihood analysis using different sets of background data combined with growth rate data $f(z)\sigma_8(z)$ for various homogeneous and clustered HDE models. The standard $\Lambda$CDM universe is shown for comparison. Values inside the parenthesis belong to clustered HDE models.   
	}
	\begin{tabular}{c  c  c c c c}
		\hline \hline
		Model  & Model 1 & Model 2  & Model 3  &$\Lambda$CDM\\
		\hline
		$k$ & 5& 4& 5& 4\\
		\hline
		$\chi^2_{\rm min}$\{total\} & 599.61 (599.34) & 739.90 (740.75) & 687.15 (688.20) & 596.08& \\
		\hline
		AIC \{total\}  & 609.61 (609.34) & 747.90 (748.75) & 697.15 (798.20) & 604.08 & \\
		\hline
		BIC \{total\}  & 630.96 (631.69) & 765.78 (766.63) & 719.50 (720.55) & 621.96 & \\
		\hline \hline 
	\end{tabular}\label{tab:best22}
\end{table*}

\begin{table*}
	\centering
	\caption{Best-fit parameters for the various homogeneous HDE models and concordance $\Lambda$CDM universe using the combined background and growth rate data.
	}
	\begin{tabular}{c  c  c c c c}
		\hline \hline
		Model  & Model(1)& Model(2)& Model(3)& $\Lambda$CDM\\
		\hline
		$\Omega_{\rm m}$ & $0.2680_{-0.0084,-0.016,-0.022}^{+0.0084,+0.017,+0.023}$&  $0.2361_{-0.0069,-0.013,-0.017}^{+0.0069,+0.014,+0.018}$ & $0.2405_{-0.0073,-0.014,-0.019}^{+0.0073,+0.014,+0.019} $ & $0.2769_{-0.0084,-0.016,-0.022}^{+0.0084,+0.017,+0.022}$ \\
		\hline
		$ H_0 $& $69.12_{-0.94,-1.8,-2.4}^{+0.94,+1.9,+2.4}$&$75.05_{-0.78,-1.5,-1.9}^{+0.78,+1.5,+1.9} $ & $73.56_{-0.79,-1.5,-1.9}^{+0.79,+1.5,+2.1} $ &  $69.74_{-0.77,-1.5,-2.0}^{+0.77,+1.5,+2.0} $&\\
		\hline
		$ \sigma_8 $& $0.771_{-0.023,-0.044,-0.057}^{+0.023,+0.044,+0.058} $&$0.866_{-0.024,-0.047,-0.063}^{+0.024,+0.047,+0.061} $ & $0.840_{-0.025,-0.050,-0.069}^{+0.025,+0.047,+0.066}$ &  $0.756_{-0.020,-0.039,-0.050}^{+0.020,+0.041,+0.054}$&\\
		\hline
		$ n $ & $0.801^{+0.049,+0.11,+0.16}_{-0.061,-0.10,-0.12}$&-- & --&--& \\
		\hline
		$ \beta$ &--&--& $0.4057_{-0.0041,-0.0080,-0.010}^{+0.0041,+0.0080,+0.010}$&--&\\ 
		\hline \hline
	\end{tabular}\label{tab:bestfit33}
\end{table*}

\begin{table*}
	\centering
	\caption{Best-fit parameters for the various clustered HDE models  using the cosmological data for background+growth analysis.
	}
	\begin{tabular}{c  c  c c c c}
		\hline \hline
		Model  & Model(1)& Model(2)& Model(3)& $\Lambda$CDM\\
		\hline
		$\Omega_{\rm m}$ & $0.2673_{-0.0083,-0.016,-0.021}^{+0.0083,+0.017,+0.023} $&  $0.2364_{-0.0071,-0.013,-0.017}^{+0.0071,+0.014,+0.019}$ & $0.2413^{+0.0067,+0.015,+0.019}_{-0.0078,-0.014,-0.018}$ & $-$ \\
		\hline
		$ H_0 $& $69.21_{-0.91,-1.7,-2.3}^{0.91,+1.8,+2.4}$&$75.03_{-0.80,-1.6,-2.0}^{+0.80,+1.6,+2.1}$ & $73.48_{-0.79,-1.6,-2.0}^{+0.79,+1.5,+2.0}$ &  $-$&\\
		\hline
		$ \sigma_8 $& $0.749_{-0.021,-0.041,-0.053}^{+0.021,+0.041,+0.057}$&$0.868_{-0.024,-0.048,-0.063}^{+0.024,+0.046,+0.059}$ & $0.839_{-0.025,-0.049,-0.060}^{+0.025,+0.048,+0.062}$ &  $-$&\\
		\hline
		$ n $ & $0.797^{+0.047,+0.10,+0.14}_{-0.054,-0.096,-0.12}$&-- & --&--& \\
		\hline
		$ \beta$ &--&--& $0.4052_{-0.0041,-0.0084,-0.010}^{+0.0041,+0.0075,+0.0097}$&--&\\
		\hline \hline
	\end{tabular}\label{tab:bestfit44}
\end{table*}

 \section{HDE models against observational data}\label{data_analysis}
 
In this section, we first implement a likelihood statistical 
analysis in order to 
place constraints on the free parameters of the current HDE models using solely
expansion data. Second, utilizing the growth rate data 
we check the 
performance of the current HDE models at the perturbation level.
Finally, based on the Akaike and Bayesian information criteria 
we study the ability of the combined (expansion+growth) data in 
constraining the cosmological parameters
of HDE models and we statistically compare them against the $\Lambda$CDM model.

\subsection{Expansion data}
 Let us start with a brief description of
the expansion data. Specifically, the latest expansion data used in our analysis are
SnIa \citep{Union2.1:2012}, BAO \citep{Beutler:2011hx,Padmanabhan:2012hf,
	Anderson:2012sa,Blake:2011en}, CMB \citep{Hinshaw:2012aka}, BBN \citep{Serra:2009yp,Burles:2000zk}, Hubble data \citep{Moresco:2012jh,Gaztanaga:2008xz,Blake:2012pj,Anderson:2013zyy}. 
In order to trace the Hubble relation we use 580 SnIa provided by the 
Union2.1 sample \citep{Union2.1:2012} and  
37 $H(z)$ measurements from the Hubble data (see Table 2). 
Moreover, we include in the analysis 
the BAO data based on 6 distinct measurements 
of the baryon acoustic scale \citep[see Tab.1 of][ and references therein]{Mehrabi:2015hva}. 
and the WMAP  
data concerning the position of CMB acoustic peak as described in \cite{Shafer:2013pxa} \citep[see also][]{Mehrabi:2015hva}.
Lastly, we utilize 
the Big Bang Nucleosynthesis (BBN) point which 
constrains the value of $\Omega_{\rm b0}$ \citep{Serra:2009yp}. 
For more details regarding the MCMC technique used, we refer the reader to \cite{Mehrabi:2015hva} \citep[see also][]{Basilakos:2009wi,Hinshaw:2012aka,Mehrabi:2015kta,Mehrabi:2016exz,Malekjani:2016edh}.

Following standard lines the overall likelihood function 
is written as the product of the individual likelihoods:
\begin{equation}\label{eq:like-tot}
{\cal L}_{\rm tot}({\bf p})={\cal L}_{\rm sn} \times {\cal L}_{\rm bao} \times {\cal L}_{\rm cmb} \times {\cal L}_{\rm h} \times
{\cal L}_{\rm bbn}\;,
\end{equation}
and thus the total chi-square $\chi^2_{\rm tot}$ is given by:
\begin{equation}\label{eq:like-tot_chi}
\chi^2_{\rm tot}({\bf p})=\chi^2_{\rm sn}+\chi^2_{\rm bao}+\chi^2_{\rm cmb}+\chi^2_{\rm h}+\chi^2_{\rm bbn}\;,
\end{equation}
where the statistical vector ${\bf p}$ includes the 
free parameters that we
want to constrain.
In our case this vector becomes:
(a) ${\bf p}=\{\Omega_{\rm DM0},\Omega_{\rm b0}, H_0, n\}$ for model (1), (b)
${\bf p}=\{\Omega_{\rm DM0},\Omega_{\rm b0}, H_0\}$ for model (2)
and (c) ${\bf p}=\{\Omega_{\rm DM0},\Omega_{\rm b0}, H_0, \beta\}$ in the case of model (3).
Notice that regarding the value of $\Omega_{\rm r0}$ we have set it 
to $\Omega_{\rm r0}=2.469\times 10^{-5}h^{-2}(1.6903)$ 
where $h=H_0/100$ \citep{Hinshaw:2012aka}.

Furthermore, in order to identify the statistical significance of our results
we utilize the well known {\rm AIC} and {\rm BIC} criteria.  
Assuming Gaussian errors the {\rm AIC} \citep{Akaike:1974} and {\rm BIC} \citep{Schwarz78} estimators  
are given by
\begin{eqnarray}
{\rm AIC} = -2 \ln {\cal L}_{\rm max}+2k+\frac{2k(k+1)}{N-k-1} \label{eq:AIC}\;,\\
{\rm BIC} = -2 \ln {\cal L}_{\rm max}+k\ln{N}\;,
\end{eqnarray}
where $N$ is the total number of data and $k$ is the number of free parameters \citep[see also][]{Liddle:2007fy}.
Our main statistical results are shown 
in Tables (\ref{tab:best}) and (\ref{tab:bestfit}),
in which we provide the goodness of fit statistics
($\chi_{\rm min}^2$, {\rm AIC}, {\rm BIC}) and the fitted cosmological parameters
with the corresponding $\sigma$ uncertainties,  
for three different HDE models (see section \ref{background-1}). For comparison we also
present the results of the concordance $\Lambda$CDM model.
From the viewpoint of {\rm AIC} analysis, it is clear that 
a smaller value of {\rm AIC} implies a better model-data fit.
Also, in order to test,
the statistical performance of the different models in reproducing
the observational data, we need to utilize the model
pair difference $\Delta {\rm AIC}={\rm AIC}_{\rm model}-{\rm AIC}_{\rm min}$.
It has been found that the restriction 
$4<\Delta {\rm AIC} <7$ suggests a positive evidence 
against  
the model with higher value of ${\rm AIC}_{\rm model}$ \citep{Burnham2002,Burnham2004}, while 
the inequality $\Delta {\rm AIC} \ge 10$ suggests a strong such evidence. 
In this framework, for $\Delta {\rm AIC} \le 2$ we have an indication 
of consistency between the two comparison models. 
Concerning the {\rm BIC} criterion, the model with the  
lowest {\rm BIC} value is the best model. The model pair difference 
$\Delta {\rm BIC}={\rm BIC}_{\rm model}-{\rm BIC}_{\rm min}$ provides 
the following situations:
(i) $\Delta {\rm BIC} \le 2$ indicates that the comparison model is 
consistent with the best model, (ii) the inequality $2<\Delta {\rm BIC} <6$ 
points positive evidence against the comparison model, while   
for $\Delta {\rm BIC} >10 $ such evidence becomes strong. 

As expected, after considering the aforementioned arguments
we find that the best model is the $\Lambda$CDM 
model and thus ${\rm AIC}_{\rm min}\equiv {\rm AIC}_{\Lambda}$.
${\rm BIC}_{\rm min}\equiv {\rm BIC}_{\Lambda}$.
We also find a strong evidence against HDE models (2) and (3), since the 
corresponding pair difference is $\Delta {\rm AIC}>10$ and $\Delta {\rm BIC} >10$. 
Moreover, we observe a relative weak evidence against the HDE model (1), $\Delta {\rm AIC} \simeq 5.6$ and $\Delta {\rm BIC} \simeq 9$. It is interesting to mention 
that our results are in agreement with the theoretical results of \citep{Basilakos:2014tha} \citep[see also][]{Xu:2016grp}.
\cite{Basilakos:2014tha} 
who proved that the HDE models (2) and (3) for which both kind of
Hubble terms $H$ and ${\dot H}$ appear
in the effective dark energy $\rho_{\rm de}$
are not viable neither at the background nor at the 
cosmic perturbations level.  
But let us try to understand the reason that 
HDE models (2) and (3) are disfavored by the expansion data.

Using the best fit values of Table (\ref{tab:bestfit}),
we show in Fig.(\ref{fig:energy_density}) 
the evolution of energy densities, namely radiation $\Omega_{\rm r}$,
pressureless matter $\Omega_{\rm m}$ and dark energy $\Omega_{\rm de}$ for the 
current HDE models. 
As far as the dark energy density is concerned 
we utilize the relations 
(\ref{eqn:difOmega_DE}, \ref{eq:omega_de_model2} \& \ref{eq:omeg_DE_model3}) 
that correspond to HDE model (1), model (2) and model (3), respectively.
On top of that we also plot the evolution of energy densities 
of the $\Lambda$CDM model for comparison. 
As expected in the case of $\Lambda$CDM model, we see that 
in the early matter dominated era the DE density is negligible
with respect to the other components. 
However, for HDE models (2) \& (3) we find that the 
DE component affects the cosmic expansion at early times. 
For example, prior to $z\simeq 1100$
we find that $\Omega_{d}$ tends to 0.18 and 0.06 for HDE models (2) and (3)   
respectively. The latter has an impact on the cosmic expansion 
and eventually it leads to the aforementioned statistical result, namely 
HDE models (2) and (3) are ruled out by the expansion data.
Therefore, for the rest of the current sub-section 
we focus on HDE model (1).
In this case we obtain $n= 0.785^{+0.042,+0.10,+0.15}_{-0.056,-0.094,-0.11} $ 
which implies that the present value of the EoS parameter of this model can 
cross the phantom line $w(z=0)\simeq -1.06$. 
Notice, that similar results can be found 
in previous works \citep{Shen:2004ck,Kao:2005xp,Yi:2006bw,Zhang:2012qra,Huang:2004wt,Zhang:2005hs,
Chang:2005ph,Zhang:2007sh,Ma2009,Xu:2012aw,Xu:2013mic,Li:2013dha,Mehrabi:2015kta}. 
In Fig. (\ref{fig:back}) we plot the $1\sigma$, $2\sigma$ and $3\sigma$ 
confidence regions in various planes for HDE and 
$\Lambda$CDM models respectively.
The negative correlation between the Hubble constant 
$H_0$ and the matter energy density  $\Omega_{\rm m}$ 
(baryons+dark matter) implies that if we increase 
the amount of the present energy density of matter then 
the Hubble constant decreases.  
In the case of HDE model (1), the negative correlation between model 
parameter $n$ and $H_0$ is also present. 
Now, utilizing the cosmological parameters of 
Table (\ref{tab:bestfit}) we plot the Hubble 
parameter $H(z)$ (left panel) and the 
distance modulus $\mu(z)=m-M$ of SNIa (right panel) 
as a function of redshift in Fig.(\ref{fig:SNIa}).  
The predictions of the concordance $\Lambda$CDM are also shown for 
comparison (solid curve). 
From both diagrams, we observe that HDE model (1) 
is relatively close to that of the standard $\Lambda$CDM model. 

\subsection{Growth rate data}\label{sect:growthdata}
 Now we focus our analysis at the perturbation level, namely 
we use only the growth rate data.
Since the growth data are given in terms of 
$f(z)\sigma_8(z)$ the first step is to theoretically calculate
the latter quantity for the HDE models. Notice, that 
$f=d\ln{\delta_m}/d\ln{a}$ 
is the growth rate of matter perturbations and $\sigma_8(z)=D(z)\sigma_8(z=0)$ 
is the r.m.s. mass fluctuations within $R=8 h^{-1}$Mpc at redshift $z$, where
$D(z)=\delta_{\rm m}(z)/\delta_{\rm m}(z=0)$ is the linear growth factor. 
Using 18 robust and independent
$f\sigma_8$ measurements ( see Table 3), we perform 
a likelihood analysis involving the growth data. 
Note, that even though the number of growth data
has increased greatly since 2010, 
the data are not independent from each other and thus they 
should not be used all together at the same 
time  \citep[ for mode details see][]{Nesseris:2017vor}. 
For the growth rate data the corresponding 
likelihood function ${\cal L}_{\rm gr}({\bf p})$ is written as
  \begin{equation}
  \chi^2_{\rm gr}({\bf p})=\sum_{i}\frac{\left([f(z_i,{\bf p})\sigma_8(z_i,{\bf p})]_{theor}-[f(z_i)\sigma_8(z_i)]_{obs}\right)^2}{\sigma_{\rm i}^2}\;,
  \end{equation}
   where the subscript "theor" indicates the theoretical 
value, "obs" stands for observational value and $\sigma_i$ is the uncertainty
   of the growth data. Here the statistical vector ${\bf p}$ 
includes
an additional free parameter, namely $\sigma_{8}$ which is 
the present value of the rms fluctuations. 
The statistical results of this section 
are summarized in Tab.(\ref{tab:bestfsigma1}).
Notice, that this table does not include the Hubble constant, since 
$H_{0}$ enters only in the radiation density parameter 
$\Omega_{\rm r0}=2.469\times 10^{-5}h^{-2}(1.6903)$ \citep{Hinshaw:2012aka}, hence
it is not really affect the cosmic expansion at relatively 
low redshifts.
Also, in Fig. \ref{fig:back+growth22} (green contours) 
we visualize $1\sigma$, $2\sigma$ and $3\sigma$ confidence 
levels in the $\sigma_{8}-\Omega_{\rm m}$ plane.
It becomes clear that the growth data can not place 
strong constraints on the cosmological models. Moreover, 
AIC and BIC tests suggest that $\Lambda$CDM is the best model. 
However, we find that using the growth data alone 
the corresponding pair differences of all 
HDE models are: $\Delta {\rm AIC} \le 2$ and $\Delta {\rm BIC} \le 3$.
The latter implies that the current HDE models are statistically 
equivalent with that of $\Lambda$CDM at the perturbation level, regards-less 
the status of the DE.

    \subsection{Combined expansion and growth data}
    In this section using the MCMC algorithm 
we implement an overall statistical analysis combining the 
expansion and the growth rate data. 
The statistical results concerning the clustered and homogeneous HDE models 
are presented in Tables (\ref{tab:bestfit33} \& \ref{tab:bestfit44}). 
Notice, that in the case of $\Lambda$CDM model DE does not cluster.
Comparing the expansion and the growth rate data 
we find that the best fit parameters are 
roughly the same (within $1\sigma$ errors) 
to those provided by the background data. 
In Figs.(\ref{fig:back+growth1} \& \ref{fig:back+growth2}), we present 
the $1\sigma$, $2\sigma$ and  $3\sigma$ confidence contours 
for different types of DE, namely homogeneous and clustered. 
Based on the joint (background+growth) 
analysis the {\rm AIC} and {\rm BIC} tests show that 
the $\Lambda$CDM model is the best model, while they 
indicate that the cosmological data disfavor 
HDE models (2 \& 3). 
Moreover, {\rm AIC} suggests weak evidence against
HDE model (1)  
($\Delta {\rm AIC} \simeq 5.5$), while 
BIC indicates strong such evidence ($\Delta {\rm BIC} \simeq 8$). 
Therefore, based on the latter comparison we cannot reject this model. 
In Fig.(\ref{fig:back+growth22}), we plot the likelihood 
contours (red plots) in the $\sigma_{8}-\Omega_{\rm m}$ plane. 
Evidently, the combined
analysis of expansion and growth rate data 
reduces significantly
the parameter space, providing tight constraints
on the cosmological parameters.  
Finally, with the aid of the best fit solutions provided 
in Tables (\ref{tab:bestfit33} \& \ref{tab:bestfit44}), we plot in Fig.(\ref{fig:fsigmahom})
the evolution of $f(z)\sigma_(z)$ for homogeneous and clustered HDE models 
respectively. Notice that the solid points correspond 
to the growth data. As we have already described above the explored 
cosmological models are in agreement 
with the growth rate data.

 \section{conclusion}\label{conclusion}
 In this work we compared the most popular Holographic 
dark energy models with  
the latest observational data. 
These HDE models were constructed on the basis of 
the event horizon IR cutoff (model 1), the Ricci scale IR cutoff (model 2) and 
the Granda \& Oliveros (GO) IR cutoff (model 3), respectively. 
Initially, we implemented 
a standard likelihood analysis using the latest 
expansion data 
(SNIa, BBN, BAO, CMB and $H(z)$) and we  
placed constraints on the free parameters of the HDE models.
Combining the well known Akaike and Bayesian
information criteria we found that the data 
disfavor the HDE models (2) and (3). We also found that the HDE model (1)  
cannot be rejected by the geometrical data.  The latter result can be understood in the context of early dark energy,
namely unlike HDE model (1), the rest of the HDE models   
predict a small but non-negligible amount of DE at early enough times.
As expected, after considering the aforementioned statistical tests
we verified that the best model is the $\Lambda$CDM 
model. Moreover, we found that the latter results
remains unaltered if we combine the growth rate 
data with those of the expansion data.
Finally, focusing at the perturbation level, namely using only 
the growth rate data we found that the current HDE models are in agreement with the data, 
regard-less the status of the DE component (homogeneous or clustered). 
However, we found that the growth rate data alone can not be used 
toward constraining the HDE models.

\bibliographystyle{mnras}
\bibliography{ref}
\label{lastpage}
	
\end{document}